\documentclass{aa}
\usepackage{graphicx}
\usepackage{hyperref}
\usepackage[varg]{txfonts}
\usepackage{tikz}
\usepackage{natbib}
\usetikzlibrary{shapes.geometric, arrows}

\tikzstyle{node} = [rectangle, text centered, text width=2cm, minimum height=1cm]
\tikzstyle{layer} = [rectangle, text centered, draw=black, minimum width=5cm]
\tikzstyle{arrow} = [->, align=center]

\bibpunct{(}{)}{;}{a}{}{,} 

\begin{document}

\title{Active deep learning method for~the~discovery of~objects of~interest in~large spectroscopic surveys\thanks{Based on spectra obtained with 2~m Perek Telescope of Ond\v{r}ejov observatory, Czech Republic and archival LAMOST DR2 spectra.}\fnmsep\thanks{Catalogues of our emission-line candidates  are available only in electronic form at the CDS via anonymous ftp to {\tt cdarc.u-strasbg.fr} (130.79.128.5) or via {\tt http://cdsweb.u-strasbg.fr/cgi-bin/qcat?J/A+A/.}}}
\author{P. Škoda\inst{1,2} \and O. Podsztavek\inst{2} \and P. Tvrdík\inst{2}}

\institute{
    Astronomical Institute of the Czech Academy of Sciences,
    Fričova 298, 25165 Ondřejov, Czech Republic \\
    \email{skoda@sunstel.asu.cas.cz}
    \and
    Faculty of Information Technology, Czech Technical University in Prague,
    Thákurova 9, 16000 Prague 6, Czech Republic \\
    \email{\{skodape4,podszond,tvrdik\}@fit.cvut.cz}
}

\date{}

\titlerunning{Active deep learning in large spectroscopic surveys}


\abstract
{
        Current archives of the LAMOST telescope contain millions of pipeline-processed spectra
        that have probably never been seen by human eyes.
        Most of the rare objects with interesting physical properties,
        however, can only be identified by visual analysis of their characteristic spectral features.
        A proper combination of interactive visualisation with modern machine learning techniques opens new ways to discover such objects.
}
{
        We apply active learning classification methods supported by deep convolutional neural networks to automatically identify complex emission-line shapes in multi-million spectra archives.
}
{
        We used the pool-based uncertainty sampling active learning method driven by a custom-designed deep convolutional neural network with 12 layers.
        The architecture of the network was inspired by VGGNet, AlexNet, and ZFNet,
        but it was adapted for operating on one-dimensional feature vectors.
        The unlabelled pool set is represented by 4.1 million spectra from the LAMOST data release 2 survey.
        The initial training of the network was performed on a labelled set of about 13\,000 spectra obtained in the 400~\AA{} wide region around H\(\alpha\) by the 2~m Perek telescope of the Ondřejov observatory, which mostly contains spectra of Be and related early-type stars.
        The differences between the Ondřejov intermediate-resolution and the LAMOST low-resolution spectrographs were compensated for by Gaussian blurring and wavelength conversion.
}
{
        After several iterations, the network was able to successfully identify emission-line stars with an error smaller than 6.5\%.
        Using the technology of the Virtual Observatory to visualise the results, we discovered 1\,013 spectra of 948 new candidates of emission-line objects
        in addition to 664 spectra of 549 objects that are listed in SIMBAD
        and 2\,644 spectra of 2\,291 objects identified in an earlier paper of a Chinese group led by Wen Hou.
        The most interesting objects with unusual spectral properties are discussed in detail.
}
{}

\keywords{methods: statistical --
    techniques: spectroscopic --
    stars: emission-line, Be --
    line: profiles --
    virtual observatory tools
}

\maketitle



\section{Introduction}

The stellar spectral classification,
as explained in \citet{gray-spec}, is an important astrophysical task of assigning a particular label
(mixture of letters and Arabic and Roman numbers),
called the spectral class,
to each spectrum based on the visual similarities
(e.g. presence, strength, and width of the spectral lines of a given element, or a combination of multiple lines).
A common automatic procedure \citep[see e.g.][Chap 13.5]{gray-spec} uses statistical matching
(mainly using \(\chi^2\) fitting)
of a given spectrum with an extensive set of template spectra
that may be either synthetic or come from a library of carefully selected stars (called spectral standards).
This method is also used in various modifications for the automatic spectral classification of large spectroscopic surveys,
such as the
Sloan Digital Sky Survey \citep[SDSS;][]{2008AJ....136.2022L}
or Large Sky Area Multi-Object Fiber Spectroscopic Telescope \citep[LAMOST;][]{pipeline, 2015AJ....150..187L}.

A problem arises in many cases when appropriate model of the spectrum is not known
and the library used for matching is not rich enough to contain unusual or new types.
In addition to this, many types of celestial objects may show complex shapes of only several prominent spectral lines
(mainly H\(\alpha\) or other Balmer and Paschen lines)
that cover only small parts of the whole spectrum.
The integral statistics then fails,
and target-tailored methods must be applied to discover such usually rare objects.
This is the case of various objects with emission lines that allow us to study a wide range of interesting physical processes.

Pre-main-sequence stars such as young stellar objects and T~Tau stars~\citep{reipurth96, kurosava06},
or hot stars with expanding envelopes or strong winds show prominent emission lines,
as do cataclysmic variables, novae, and even late-type stars with chromospheric activity.
See~\citet{kogure2007} or \citet{traven2015} for a comprehensive overview of these cases.

The classical Be stars~\citep{2003PASP..115.1153P}
and the rare class of B[e] stars~\citep{2003A&A...408..257Z} are other cases of well-studied objects with complicated emission-line profiles that
often look like symmetric or slightly asymmetric double peaks, sometimes superimposed on absorption lines, depending on their disk geometry \citep{2010ApJS..187..228S}.
The visual classification of their profiles~\citep{1988A&A...189..147H} is a challenging task even on small samples,
but it becomes impossible in surveys with millions of spectra.

The classical approach to finding emission lines is to compute integral statistics around their expected positions.
It is similar to the standard method of measuring the line equivalent width~\citep{2012MNRAS.425.3162K, 2013PASP..125.1164W}.

Such an integral measure based on three-pixel statistics was taken by~\citet{Lin2015} on the LAMOST data release 1 (DR1)
in order to find strong uprising peaks.
This resulted in a catalogue of 203 emission-line stars,
 23 of which were identified as classical Be stars and 180 are claimed to be discovered candidates.
In order to find double-peak profiles hidden in deep absorption,
\citet{Hou} (hereafter H16) used a more advanced method
based on the difference of several statistics with different kernel width.
The authors made an extensive analysis of the LAMOST data release 2 (DR2) survey
and published a catalogue of 11\,204 spectra of emission-line
stars\footnote{\url{http://paperdata.china-vo.org/vac/dr2/HouEmission2016.tar.gz}}.

We propose an alternative approach for the discovery of emission-line spectra
here based on machine learning of individual shapes of prominent spectral lines.
For the sake of simplicity, we limit ourselves to the vicinity of the H\(\alpha\) line.
The early attempts on a small sample of good spectra~\citep{2012ASPC..461..573S, IJAC} have already justified this method,
and its application to the LAMOST DR1~\citep{2015ASPC..495...87S,AI2016} has resulted in the discovery of unknown emission-line candidates.
This article describes the first systematic investigation of the LAMOST DR2
using a deep convolutional neural network (CNN) in combination with active learning.

We organised this article as follows.
Section~\ref{method} describes our active learning method based on CNNs in detail.
Section~\ref{experiments} shows the application of the developed method to the discovery of emission-line spectra in the LAMOST DR2.
Section~\ref{results} discusses the outcomes of the experiment
and lists examples of discovered objects of interest.
Finally, we conclude in Sect.~\ref{conclusion}.
Furthermore, we compare our method to the non-active learning scenario in Appendix~\ref{appendix:non-active}, and we provide a detailed analysis of the results in Appendix~\ref{appendix:reconfirmation}.



\section{Active deep learning method}
\label{method}

The discovery of objects of interest in large archives of astronomical spectra would be a standard machine learning task
if a large and representative labelled data sample of a given archive were available.
With such a training set, it would be straightforward to train a supervised learning model
and classify the whole archive with high accuracy.
However, our experiment in Sect.~\ref{experiments} has shown
that if there is no proper training dataset,
standard machine learning methods provide poor results with a high rate of both false and missed candidates.

This means that if the training labelled data are not a sufficiently large representation of a spectral archive,
for example, when the training set is biased or comes from another, but similar archive, other machine learning approaches need to be developed to obtain reasonable discovery results.
We propose and evaluate here an extension of a deep CNN classification method with class balancing and active learning.

The following subsections explain in detail why and how we combined a CNN with a class balancing algorithm and an active learning method.
This unified active deep learning workflow allowed us to discover objects of interest
(objects with emission-line spectra)
in the LAMOST DR2 altough only a small number of training data were available from a different spectral archive.

\begin{figure*}
    \centering
    \begin{tikzpicture}[node distance=4cm]
        \node (start) [node] {initialisation with labelled samples};
        \node (training_set) [node, right of=start] {gathered training set};
        \node (balancing) [node, right of=training_set] {balanced training set};
        \node (model) [node, right of=balancing] {trained CNN};
        \node (classification) [node, below of=model] {classified unlabelled samples};
        \node (performance) [node, left of=classification] {stopping criterion};
        \node (sampling) [node, left of=performance] {unlabelled batch of samples};
        \node (labelling) [node, left of=sampling] {labelled batch of samples};
        \node (prediction) [node, below of=performance] {predicted candidates};
        \node (candidates) [node, left of=prediction] {all candidates};
        \draw [arrow] (start) -- (training_set);
        \draw [arrow] (training_set) -- node[anchor=south] {class\\balancing} (balancing);
        \draw [arrow] (balancing) -- node[anchor=south] {training} (model);
        \draw [arrow] (model) -- node[anchor=east] {classification\\of unlabelled samples} (classification);
        \draw [arrow] (classification) -- node[anchor=south] {performance\\estimation} (performance);
        \draw [arrow] (performance) -- node[anchor=south] {uncertainty\\sampling} (sampling);
        \draw [arrow] (sampling) -- node[anchor=south] {oracle\\labels} (labelling);
        \draw [arrow] (labelling) -- node[anchor=east] {add the labelled batch\\to the training set} (training_set);
        \draw [arrow] (performance) -- node[anchor=east] {take out samples\\labelled as target} (prediction);
        \draw [arrow] (prediction) -- node[anchor=south] {add candidates\\of the oracle} (candidates);
    \end{tikzpicture}
    \caption{
        Flowchart of our active learning method.
        First, the algorithm is initialised only with labelled training data,
        the CNN is trained,
        and unlabelled spectra are classified.
        Then, uncertainty sampling selects the spectra that the network is least certain for.
        Finally, these spectra are labelled by an oracle,
        are added to the training set,
        and a new training iteration starts.
        When the performance is satisfactory,
        samples classified into target classes are taken as candidates
        and are extended with samples classified into target classes by the oracle.
    }
    \label{flowchart}
\end{figure*}
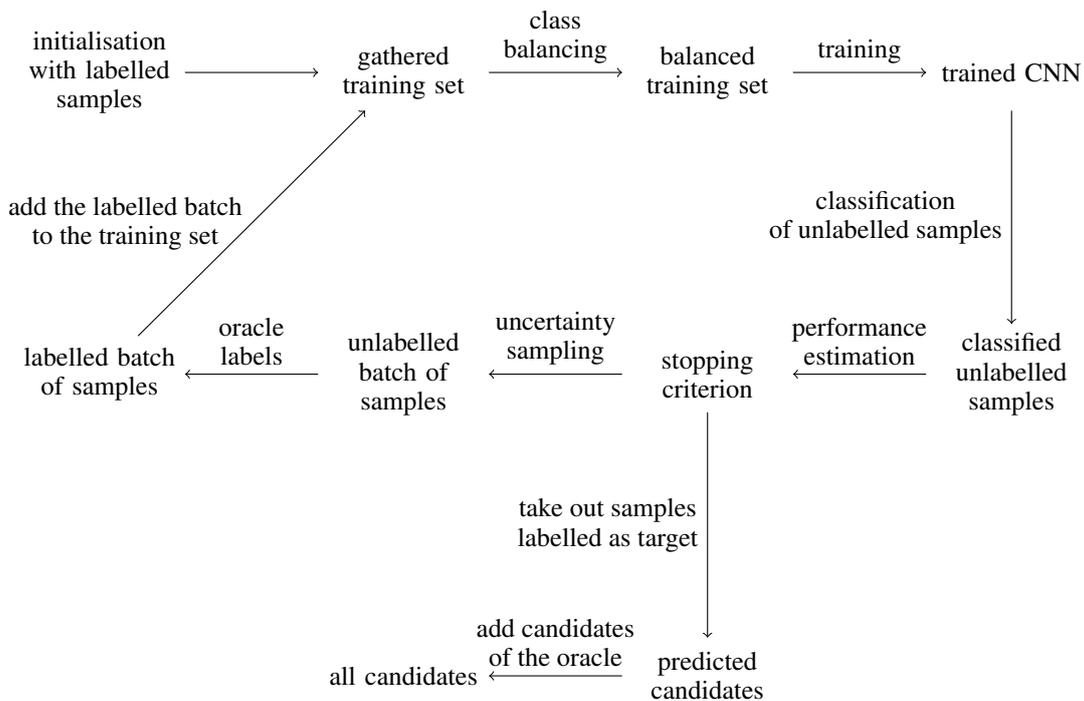

\subsection{Deep learning}

Deep learning is a type of machine learning
that solves the problem of representational learning by learning a hierarchy of concepts.
In representational learning, we try to learn a representation of the data
that would facilitate the subsequent learning task.
Deep learning allows computers to learn a good data representation
by building complicated representations out of more simple ones
\citep{lecun2015, Goodfellow-et-al-2016}.

Today, CNNs~\citep{lecun1989} are the most advanced deep learning method.
CNNs started to be recognized
when \citet{krizhevsky2012} achieved a winning top-5 test error rate of 15.3\% on the ImageNet Large Scale Visual Recognition Challenge 2012~\citep{ilsvrc}.
We wish to take advantage of CNNs because
spectra of stellar objects can be viewed as one-dimensional arrays with a single channel,
whereas a typical image is a two-dimensional array with usually three RGB channels.

The CNNs are specialised neural networks
that use convolution to process data with a grid-like topology.
A convolution leverages three essential properties of these biologically inspired networks:
sparse interactions
(kernels used for convolution with an image have fewer parameters than a fully connected layer),
parameter sharing
(rather than learning a separate set of parameters, CNNs learn one set for all locations),
and equivariance to translation
(if an object shifts in the input, its corresponding output shifts by the same distance).
Furthermore, a typical CNN has pooling layers that follow the convolution and activation layers.
The pooling layers make the representation invariant to small translations and rotations of the input.
This invariance is a useful property for application to spectra
because spectral lines might be blue- or red-shifted due to the high radial velocity
\citep{Goodfellow-et-al-2016}.

Deep CNNs have already been successfully applied in astronomy and astrophysics.
For example, \citet{aniyan2017}, \citet{sanchez2018}, and \citet{alhassan2018} used CNNs to automate the morphological classification of radio sources.
\citet{alger2018} localised host galaxies for a given radio component with a CNN
using data from experts and crowdsourced training data.
Furthermore, \citet{george2018} applied two CNN time-series data
to the detection and parameter estimation of gravitational waves from binary black hole mergers.
The two CNNs achieved a similar performance as previous advanced methods
but were much faster,
thus allowing real-time processing.
For all these reasons, we decided to develop an active deep learning method with CNNs
for the discovery of objects of interest.

\subsection{Class balancing}

When discovering rare objects of interests in large spectroscopic surveys,
we face the class imbalance problem~\citep{prati2009}.
Labelled spectra of rare objects of interest
(hereafter target spectra)
will usually be in the minority, in contrast to the labelled spectra of abundant objects
(hereafter non-target spectra).
Therefore, the labelled training data will tend to be imbalanced.
Moreover, target spectra will be in a significant minority in
general massive spectral archives (e.g. LAMOST or SDSS).

Our application of the active deep learning, see Sect.~\ref{experiments} for details, revealed exactly the class imbalance problem.
The archive of the Ondřejov 2~m Perek telescope is focused on the observation of emission-line stars.
Although there is almost the same percentage of single peaks as absorptions,
double peaks are still in the minority (see Sect.~\ref{classification}).
Moreover, there are (at least by order of magnitude) fewer emission-line spectra than standard ones in the LAMOST survey
because emission-line objects are rare in the Universe.
In these cases, class balancing has shown to be an essential part of workflows and leads to successful performance
(e.g. for the necessity of class balancing in astronomy, see \citet{calleja2011} or \citet{lyon2016},
and in medicine, see~\citet{rastgoo2016}).

To overcome the fact that CNNs will tend to discriminate the minority classes,
we incorporated in our experiments the synthetic minority over-sampling technique (SMOTE) proposed by~\citet{smote}. This technique allows enlarging the number of labelled target spectra
to the same size as the more abundant non-target spectra.

\subsection{Active learning}
\label{al}

Our experiments have shown
that the combination of a CNN and class balancing is still not sufficient for the discovery of objects of interest
because the first prediction of candidates delivered a considerable amount of false candidates and featureless noisy spectra.
The reason for this failure was an imperfect training dataset.
Therefore, we decided to explore active learning~\citep{settles09}
to circumvent the requirement of good representativeness of labelled samples
to exploit the full potential of deep neural networks to discover objects of interest.

Active learning has already shown to be successful in astronomy,  for example,
in estimating parameters of stellar population synthesis models
by~\citet{solorio05} or for the classification of light curves of variable stars
by~\citet{richards12}. \citet{gupta2016} used active learning to learn a model
for photometric data classification from spectroscopic data (the work
was extended by~\citet{vilalta2019}), and recently, active learning was used to
minimise the number of required spectroscopically confirmed labels in
preparing training sets for the photometric classification of supernova light
curves by~\citet{ishida2019a} and for active anomaly detection in light curves
of supernovae by~\citet{ishida2019b}.  Moreover, active deep learning has been successfully tested in remote sensing by~\citet{liu2017}, with further
examples reviewed in~\citet{yang2018}.  To the best of our knowledge, our
method represents the first astronomical application of active deep learning.

Active learning is a machine learning technique based on the idea
that an algorithm will perform better with fewer training data
if it is allowed to choose data for its training.
A machine learning algorithm combined with active learning (an active learner) queries unlabelled data examples
to be labelled by an oracle (e.g. a human expert).

In the case of large spectra archives,
there are huge pools of unlabelled data that can be processed and gathered at once
(a so-called pool-based setting in the context of active learning).
Spectra are queried selectively from the pool according to an informativeness measure
that evaluates all spectra in the pool.
Concerning CNNs, the most straightforward approach is to use uncertainty sampling as the query strategy.
This strategy selects spectra for which the CNN provided the least certain labelling
because the last layer of the CNN is usually a softmax layer.
This layer produces probabilities of classes for each spectrum.
Therefore, to query spectra for labelling,
we compute the information entropy,

\begin{equation}
    H = -\sum_i p_i \ln p_i,
\end{equation}

where \(p_i\) is the probability of class \(i\),
for all the spectra in the pool.
Then, the method selects spectra with the highest information entropy.

Because the training of a CNN can be time-consuming,
our method uses so-called batch-mode active learning,
which iterates in cycles:
an oracle labels a batch of queried samples in each iteration in order to save time and computational resources
(training of a CNN).
More specifically, the method selects a batch of a previously specified size
(e.g. one hundred as in our experiments in Sect.~\ref{application})
from all spectra in the pool,
and the oracle visually classifies them.
Then, we add all the visually labelled spectra to the training set,
so that it contains training data from the previous iterations and newly classified spectra.

Lastly, to decide when to stop the active learning iterative procedure,
we need to track the performance of the CNN.
The obvious possibility is to estimate a performance measure
and stop learning when a plateau is reached
(e.g. when adding newly labelled spectra would not increase the performance measure of the CNN).

When a large pool of unlabelled samples
contains a negligible number of target spectra,
it is reasonable to estimate precision, defined as

\begin{equation}
        \mathit{precision} = \frac{\mathit{TP}}{\mathit{TP} + \mathit{FP}}
,\end{equation}

where \(\mathit{TP}\) (true positive) is the number of correctly predicted target spectra,
and \(\mathit{FP}\) (false positive) is the number of incorrectly predicted target spectra.
In the case of precision, we can expect
that a random sample of spectra classified into target classes
will contain the true target spectra.
On the other hand, a random sample of all spectra or non-target spectra will probably contain only non-target spectra.
Therefore, an estimation of any performance metric based on such random samples
will not yield a useful result.
For example, an estimate of accuracy,
which has to be based on a random sample of all spectra, will almost certainly be 1 or very close to it.
Moreover, when discovering rare objects,
we are not interested in accuracy,
but rather in precision and
recall\footnote{Recall is the ratio of correctly predicted target spectra and all target spectra.}.
However, the estimation of recall faces the same problem
as the estimation of accuracy.

For this reason, we cannot have any randomly sampled performance estimation set
fixed for all iterations.
However, we have to sample a new random sample in every iteration
as the set of predicted target spectra is changing.

In summary, our active deep learning method takes the labelled data as the initial training set and balances it.
Having a balanced training set, we train the CNN
and use the trained CNN to classify all the unlabelled pool of spectra.
Then, we use the uncertainty sampling query strategy to obtain a batch of samples for labelling by an oracle
that labels all the samples in the batch.
The labelled samples are taken out of the unlabelled pool
and placed into the labelled training set.
Now, we repeat these steps
until the performance of our CNN is satisfactory.
When we are satisfied with the CNN performance,
the unlabelled samples that were lastly predicted as target ones
become new candidates of emission-line stars.
Finally, we move the samples labelled by the oracle as target
from the training set to the candidate set.
The flowchart in~Fig.~\ref{flowchart} illustrates the whole algorithm of our active deep learning method.



\section{Experiments}
\label{experiments}

To illustrate the application of our active deep learning method,
we have performed experiments with the discovery of objects with signatures of H\(\alpha\) emission in the LAMOST DR2 survey
using labelled data from the Ondřejov 2~m Perek telescope.
The following sections describe the data, the data preparation, the classes of interest, and our method application.

\subsection{Input data}
\label{data}

The archive of spectra obtained with 700~mm camera in the Coud\'e spectrograph of the 2~m Perek telescope at the Ondřejov observatory of the Astronomical Institute of the Czech Academy of Sciences is a unique source of spectra of emission-line stars
(mostly Be and B[e] stars, stars with strong winds and several novae).
This continuously growing archive (hereafter CCD700), currently contains about 17\,000 spectra,
the majority of which (more than 13\,000) are exposed in spectral range 6\,250--6\,700~\AA{} with a spectral resolving power of about 13\,000.
The standard IRAF procedure~\citep{iraf} reduces the spectra,
including the calibration in air wavelengths and heliocentric correction.

The LAMOST telescope has delivered one of the currently largest collections of optical spectra.
Four thousand fibres positioned by micro-motors feed 16 LAMOST spectrographs.
Its publicly available DR2 contains over four~million spectra with a spectral resolution power of about 1\,800, covering the range 3\,690--9\,100~\AA{}~\citep{DR2}.
The LAMOST pipeline~\citep{pipeline}  automatically assigns an estimated spectral class to the spectra.
However, the pipeline uses classification mostly based on the global shape and integral properties of a spectrum
in given band-passes using a set of predefined templates.
The local features (e.g. detailed line profiles) are ignored.
Strong narrow emissions can even be rejected by the pipeline as possibly spoiled pixels.
Therefore, we did not use the assigned spectral classes.
Hereafter we call the set of all unlabelled LAMOST DR2 spectra the LAMOST pool.
The spectral axis of the FITS files in the LAMOST archive are expressed in the logarithm of the vacuum wavelength.

\subsection{Data preprocessing}

A common assumption in machine learning is that training data (in our work, the CCD700 data) and the data of interest (the LAMOST pool) are from the same probability distribution~\citep{transfer2010}.
However, in this work, we are interested in the classification of the LAMOST pool using the training set from the Ondřejov spectrograph,
which contains mostly emission spectra. This means that the training set is highly biased.
The distribution mismatch between the training data and the data of interest is a well-known problem in machine learning and is called domain adaptation~\citep{glorot2011}.

Using the technology of the Virtual Observatory (see Appendix~\ref{appendix:votools}) for cross-matching,
we have identified only 22 spectra that were observed both by the Ondřejov 2~m Perek Telescope and by LAMOST.
Only a few (e.g. BT~CMi, HD~53\,416, or V395~Aur) of them show emission lines.
The lack of labelled training spectra in the LAMOST pool prevents the usage of supervised training.
To use the CCD700 spectra as our training set,
we therefore applied a domain transfer to the CCD700 spectra (based on optical engineering procedures),
so that they will look as if they were exposed with the LAMOST spectrograph.
\citet{dtn} claimed that domain transfer is useful when solving the domain adaptation problem.

Firstly, we applied air-to-vacuum wavelength conversion to the CCD700 spectra using formulas provided in~\citet{air2vacuum}
because spectra from the CCD700 archive are in air wavelengths,
but the LAMOST spectra use vacuum wavelengths.
Additionally, we converted the vacuum wavelengths of spectra from the LAMOST pool from the logarithmic into linear scale.

\begin{figure}
    \resizebox{\hsize}{!}{
        \includegraphics{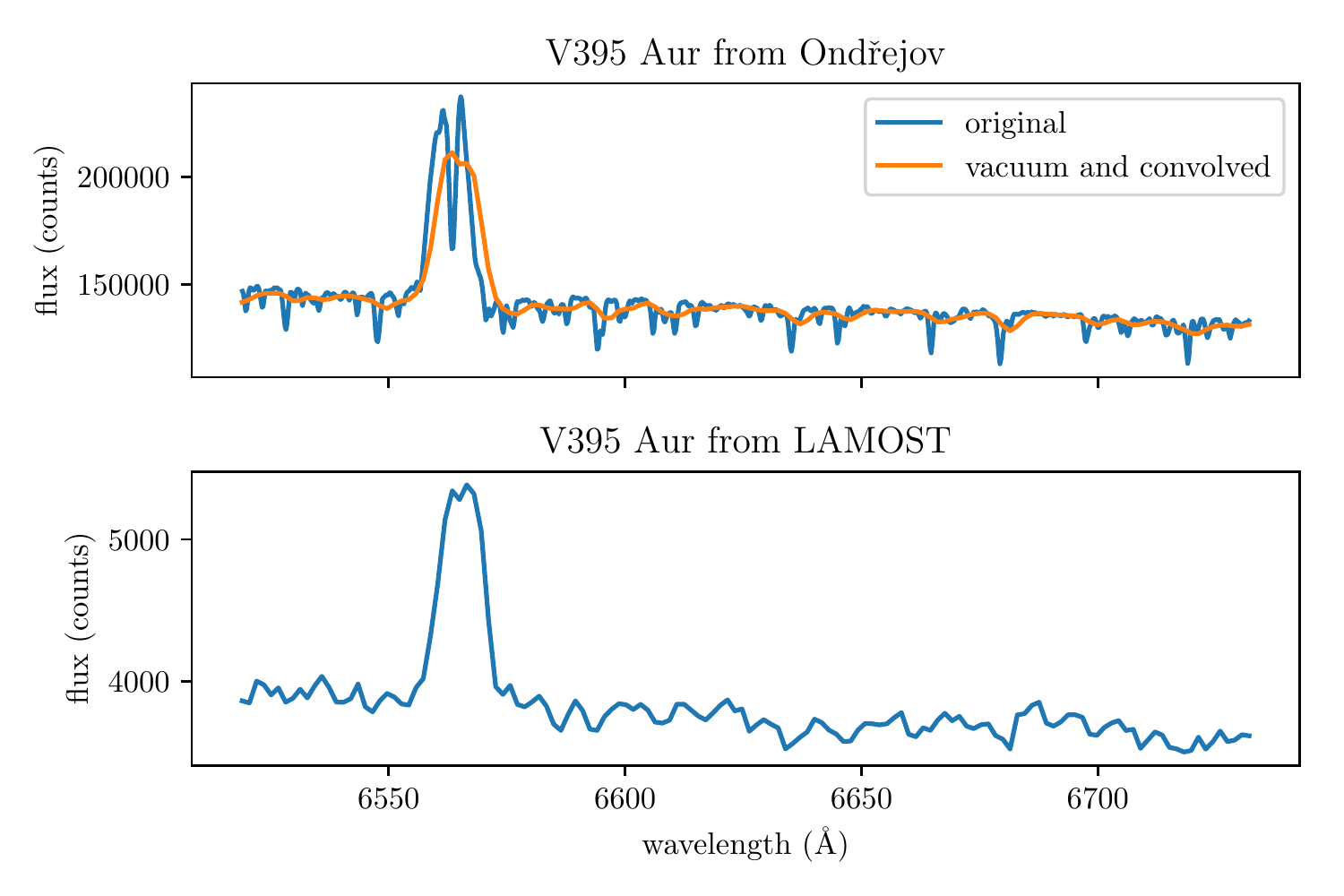}}
    \caption{
        Comparison of a LAMOST spectrum with a Ondřejov CCD700 spectrum converted into the LAMOST lower resolution and vacuum wavelengths.
    }
    \label{data_conversion}
\end{figure}

Secondly, because the CCD700 spectra have a higher spectral resolution than the LAMOST spectra,
we applied the spectral resolving power degradation to the CCD700 spectra,
roughly approximated by the convolution with the Gaussian kernel of a given pixel width to reduce the high-resolution details.
Comparison figures of simulated spectra from CCD700 and the LAMOST pool of all 22 objects mentioned above showed
that the standard deviation of seven-pixel value works best.
Figure~\ref{data_conversion} shows the comparison of an Ondřejov spectrum, a cross-matched LAMOST spectrum, and the preprocessed spectrum.

Next, the CNN requires a vector of features as an input.
To have the same features for all spectra, they need to be resampled to obtain the measurement in the same wavelengths across all spectra.
We decided to use a linear interpolation
(using the linear interpolation function of the NumPy library)
to 140 uniformly distributed wavelength points in the spectral range between 6\,519 and 6\,732~\AA{}.
We used this number of points because the LAMOST spectra mostly have this number of measurements in the given range.
We derived the range from the fact that our classification is based on the H\(\alpha\) line
and most of the CCD700 spectra are exposed between these wavelengths.
This range also contains \ion{He}{I} 6\,678~\AA{} line, which is important in Be stars.
Having resampled all spectra in the same wavelength points, we can create a design matrix required for learning,
where rows are 140-dimensional feature vectors of spectra and columns contain fluxes in specified wavelengths.

The last step of preprocessing is the min-max normalisation of the spectral flux into a unit-less range \([-1, 1]\) using the equation

\begin{equation}
    \vec{x}' = 2 \frac{\vec{x} - \min(\vec{x})}{\max(\vec{x}) - \min(\vec{x})} - 1,
\end{equation}

where \(\vec{x}\) is an input not-scaled spectrum, and \(\vec{x}'\) is a scaled spectrum.
Thus, each spectrum has a maximum flux of value 1 and a minimum of value \(-1\).
We applied this preprocessing procedure for two reasons:
we would like to classify the spectra according to their shapes
(this procedure effectively suppresses the differences in intensities),
and it obtains the value in the comfortable small-valued range that is suitable for a neural network training
(this is not a feature scaling, but a scaling across each spectrum).

\subsection{Classification}
\label{classification}

In the next step,
the preprocessed CCD700 spectra were classified by~\citet{podsztavek} according the visual shape of the H\(\alpha\) into three classes:
single peak, double peak, and absorption.
The labelled spectra resulted in a dataset of 12\,936 labelled spectra
(hereafter the Ondřejov dataset\footnote{\url{https://doi.org/10.5281/zenodo.2640970}})
that were suitable for machine learning.
The counts of spectra in classes are the following:

\begin{itemize}
    \item single peak: 5\,301 spectra (40.98\%),
    \item double peak: 1\,533 spectra (11.85\%), and
    \item absorption: 6\,102 spectra (47.17\%).
\end{itemize}

\begin{figure}
    \resizebox{\hsize}{!}{
        \includegraphics{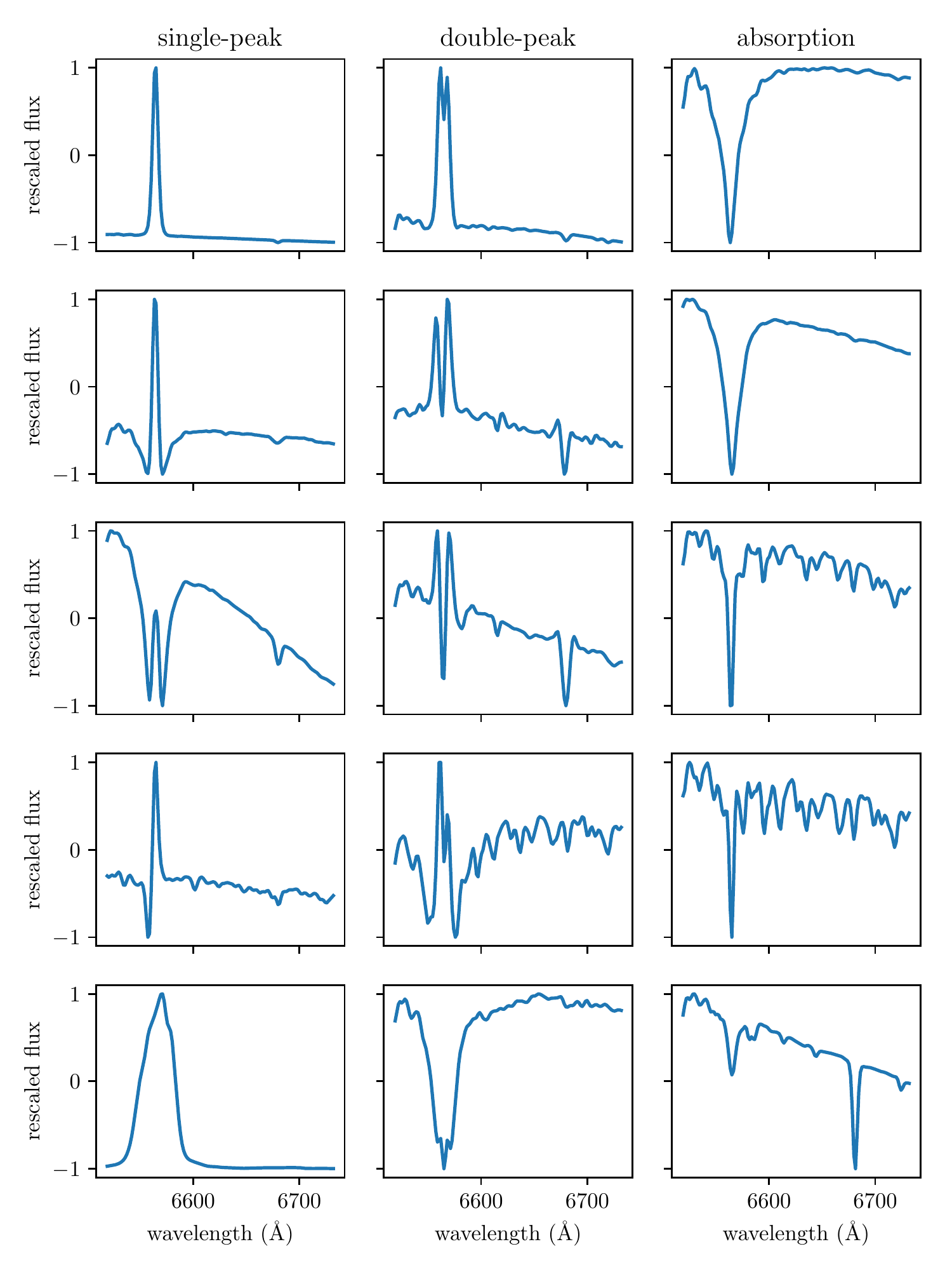}}
    \caption{Examples of spectra from all three classes in the Ondřejov dataset.}
    \label{profiles}
\end{figure}

Figure~\ref{profiles} displays representatives of each class.
In both single-peak and double-peak spectra the H\(\alpha\) line is in emission,
and the difference between the two classes is in the number of peaks, which are clearly visible in the spectrum.
Spectra in the single-peak and double-peak target classes are the target emission spectra of our interest,
and as expected, their number is smaller than the number of non-target absorption spectra, which are not interesting for us.

The Ondřejov dataset contains only well-exposed spectra,
while the LAMOST pool contains many noisy spectra with instrumental and reduction artefacts, spectra without peaks or absorption, and spectra with low signal-to-noise ratio.
During our experiment, we placed all these spectra into the non-target uninteresting class.
Therefore the non-target not interesting class contains 
bad and absorption spectra, which are both uninteresting for us.

\subsection{Application of active deep learning}
\label{application}

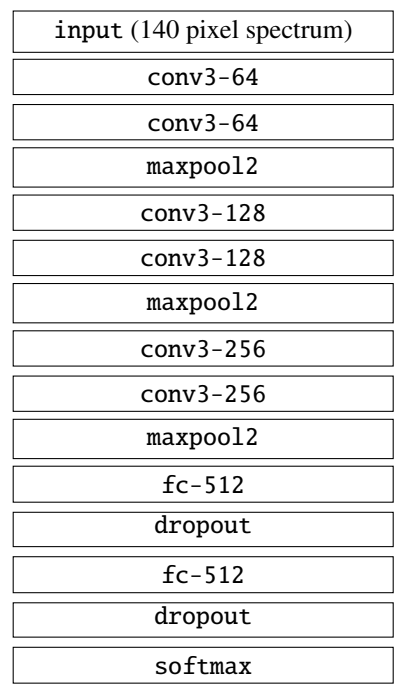
\begin{figure}
    \centering
    \begin{tikzpicture}[node distance=0.6cm]
        \node (input) [layer] {\texttt{input} (140 pixel spectrum)};
        \node (conv1) [layer, below of=input] {\texttt{conv3-64}};
        \node (conv2) [layer, below of=conv1] {\texttt{conv3-64}};
        \node (pool1) [layer, below of=conv2] {\texttt{maxpool2}};
        \node (conv3) [layer, below of=pool1] {\texttt{conv3-128}};
        \node (conv4) [layer, below of=conv3] {\texttt{conv3-128}};
        \node (pool2) [layer, below of=conv4] {\texttt{maxpool2}};
        \node (conv5) [layer, below of=pool2] {\texttt{conv3-256}};
        \node (conv6) [layer, below of=conv5] {\texttt{conv3-256}};
        \node (pool3) [layer, below of=conv6] {\texttt{maxpool2}};
        \node (fc1) [layer, below of=pool3] {\texttt{fc-512}};
        \node (dropout1) [layer, below of=fc1] {\texttt{dropout}};
        \node (fc2) [layer, below of=dropout1] {\texttt{fc-512}};
        \node (dropout2) [layer, below of=fc2] {\texttt{dropout}};
        \node (softmax) [layer, below of=dropout2] {\texttt{softmax}};
    \end{tikzpicture}
    \caption{Architecture of our CNN.
        Convolutional layers are marked as \texttt{conv3},
        where the number 3 means the size of the filter in pixels.
        The mark is followed by a dash
        and a number that specifies the count of filters.
        \texttt{maxpool2} are max-pooling layers with pool size 2, stride 2, and no padding.
        \texttt{fc-512} denotes a fully connected layer with 512 units,
        and \texttt{softmax} is the output layer.
        Dropout layers~\citep{dropout} with the hyperparameter set to the value 0.5 are used as regularisers.}
    \label{convnet}
\end{figure}

When the data were ready,
we applied our method.
We chose the architecture of a CNN as developed in previous work
that proved to be working well \citep[see][]{podsztavek}.
This CNN architecture was inspired primarily by VGGNet~\citep{vggnet}, AlexNet~\citep{krizhevsky2012}, and ZFNet~\citep{zfnet}.
However, these CNNs were designed to process multi-channel two-dimensional images.
We therefore adapted the architecture for our one-dimensional data
(replace two-dimensional convolutions with one-dimensional convolutions).
After several experiments, we converged to the architecture shown in Fig.~\ref{convnet}.
This CNN was implemented using TensorFlow~\citep{tensorflow} through the
Keras~\citep{keras} high-level interface and was run on an NVIDIA GTX980 GPU (4 GB memory, 2\,048 CUDA cores).
The network was trained with the Adam optimiser~\citep{adam} in the default setting of Keras.
The best-found weights were restored at the end of each training.
We stopped the training when the categorical cross-entropy loss function was not improved by at least \(10^{-4}\) during the last ten iterations.

After we trained the CNN with the Ondřejov dataset (the initial training set) balanced with SMOTE,
we used the model to predict classes and probabilities of classes for all spectra in the LAMOST pool.
From all the classified spectra,
a batch of 100 spectra with the highest information entropy computed from the probabilities of classes was selected
(the uncertainty sampling strategy),
visually reviewed by us (in the role of the oracle), and classified.
Then, all the 100 visually labelled spectra were moved to the training set
and removed from the LAMOST pool.
Hence, after the first iteration, the training set contained the spectra from the Ondřejov dataset and 100 new spectra from the LAMOST pool.

To track the performance of our CNN,
we decided to estimate the precision
(the ratio of correctly predicted single-peak and double-peak spectra in all predicted target spectra)
in each iteration
because of the reasons stated in~Sect.~\ref{al}.
Therefore we randomly selected 30 spectra classified into
single-peak and double-peak (target spectra) classes from the LAMOST pool
(hereafter the performance estimation sample).
The size of 30 was chosen as a good trade-off between confidence and the demands of visual verification.
Then we manually labelled the performance estimation sample
and compared our labels with the labels predicted by our CNN.
Thus we estimated the precision after each iteration.
The performance estimation sample of 30 spectra functions as a test set.
In a standard machine learning scenario, a test set is a random sample of all unseen data
that could be put into the CNN.
In our case, all possible data for our CNN are in the so far unlabelled LAMOST pool.
Therefore, the performance estimation sample will provide an unbiased estimate of precision.
We would like to point out that the manual labelling of the performance estimation sample is different from the labelling by the oracle.
The labels of the performance estimation sample are forgotten after the precision estimation,
and the spectra are left in the LAMOST pool.

\begin{figure}
    \resizebox{\hsize}{!}{\includegraphics{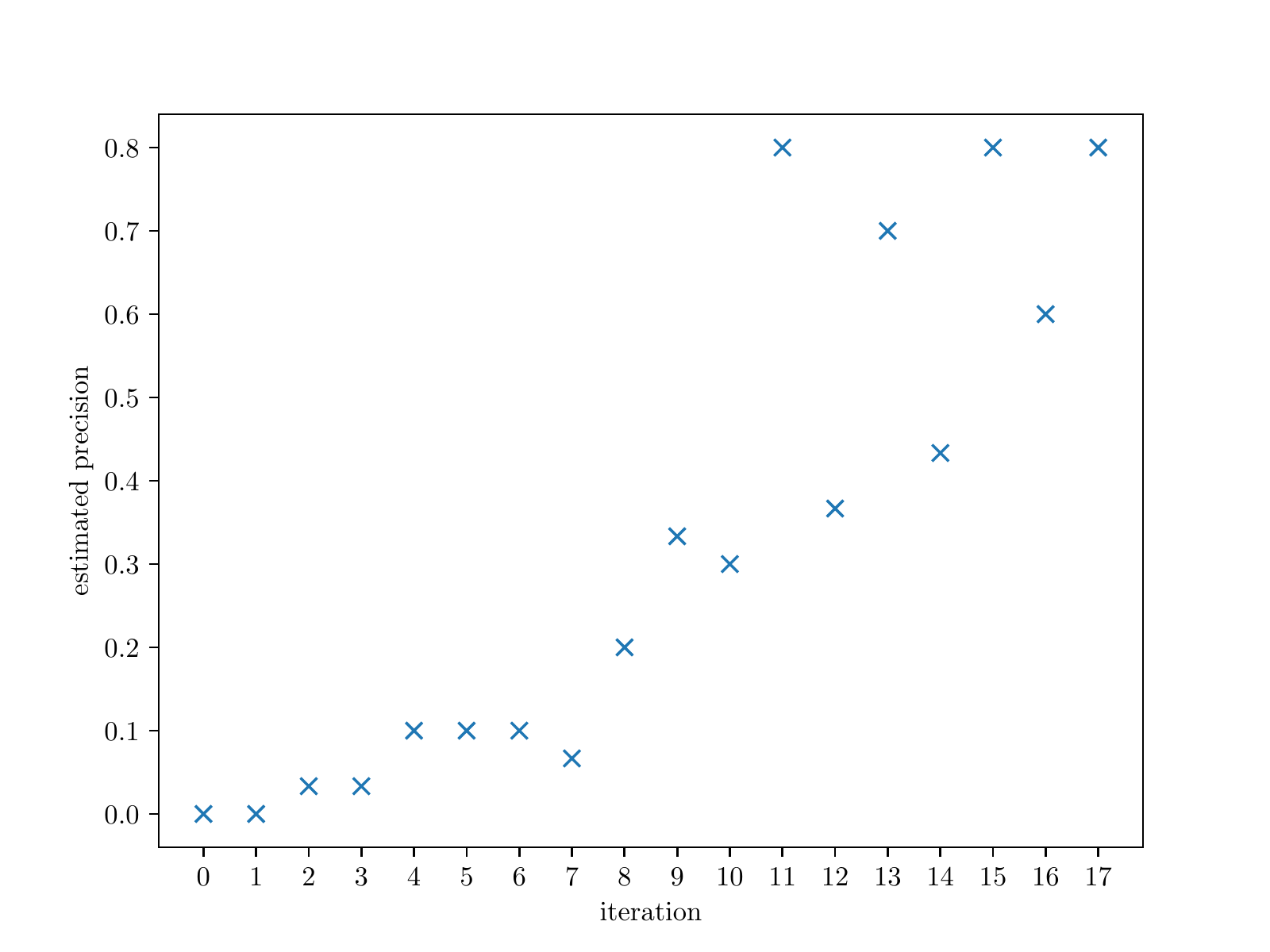}}
    \caption{Estimated precision from a sample of 30 single-peak and double-peak spectra for each iteration
        (the zeroth iteration is estimated when the CNN is trained only with the initial Ondřejov dataset).}
    \label{performance}
\end{figure}

Finally, we stopped our experiment in the 17th iteration when the estimated precision
reached more than the predefined threshold
(in our case 80\%) for the third time.
We chose the values of these parameters as a trade-off between time and performance requirements, and it can be chosen differently for different datasets.
Figure~\ref{performance} displays the precision of our CNN over 17 active learning iterations.

Because the training of our CNN was time-consuming,
we sped up the method by training the CNN during the active learning phase for a smaller number of epochs.
Then, after the active learning phase,
we ran the Adam optimisation algorithm of the CNN for a longer time
(the training was stopped when the loss function did not improve by \(10^{-5}\) during 100 training iterations)
to ensure that good convergence was achieved,
and thus fewer false candidates will be produced.
In the following text, we refer to this step as~long training.

\section{Results}
\label{results}

Our method identified 4\,379 candidate spectra with signatures of emission-line profiles
including candidates found by the oracle
in all the 4\,136\,482 LAMOST DR2 spectra.
The last CNN predicted 3\,574 spectra as single-peak and 587 as double-peak profiles,
while the oracle found 157 single-peak candidates and 61 double-peak candidates during the active learning phase.
As explained earlier, it also includes absorption profiles with small visible disturbances
that may be caused by additional circumstellar emissions.
After visual inspection (see Appendix~\ref{appendix:reconfirmation}) of the predicted candidates,
we rejected 58 as bad
(partly destroyed, noisy, or with pure absorption profiles)
and computed the partial confusion matrix in~Table~\ref{confusion}.
Finally, we had a set of 4\,321 spectra
of about\footnote{The exact number of individual objects is difficult to estimate because of cross-matching problems, as explained in Appendix~\ref{appendix:multiplicity}.} 3\,788 individual objects.

\begin{table}
    \caption{
        Partial confusion matrix of the final classification of our experiment
        (excluding candidates found by the oracle).
        The numbers show the percentage and counts (in brackets)
        of correctly predicted spectra of all spectra
        predicted for a given class.
        The 4161 spectra in this table are all the candidates predicted as single or double peaks after the long training.
        After we visually reviewed all of them,
        we found that 58 of candidates are uninteresting spectra
        (37 predicted as single peaks and 21 predicted as double peaks).
        The target classes also include some misclassification: 53 double peaks are classified as single peaks,
        and 18 single peaks are classified as double peaks.
        We note that we were unable to compute the last row of the uninteresting class
        because it would mean visually classifying all the four million spectra that are predicted as uninteresting.
    }
    \label{confusion}
    \centering
    \begin{tabular}{c|ccc}
        \hline\hline
        Predicted & \multicolumn{3}{c}{Actual class} \\
        class & single peak & double peak & uninteresting \\
        \hline
            single peak & 97.5\% (3\,484) & 1.5\% (53) & 1.0\% (37) \\
            double peak & 3.1\% (18) & 93.4\% (548) & 3.6\% (21) \\
        \hline
    \end{tabular}
\end{table}

This set includes 2\,644 spectra of 2\,291 objects that have been found previously by H16,
and 664 new spectra of 549 objects that are listed in SIMBAD
(which were not found by H16).
Our method proved to be reliable (with an error smaller than 6.5\%) because most of the candidates are classified in SIMBAD as various cases of emission-line objects,
such as cataclysmic variables, young stellar objects, dwarf novae, symbiotic binaries, IR excess objects from IRAS, classic Be stars and HAEBe stars.
In addition, our method found 1\,013 spectra of 948 new objects that are neither known in SIMBAD
nor discovered by H16.

The newly discovered objects span almost all spectral classes
as assigned by the LAMOST stellar pipeline,
but also many unclassified ones.
The visual inspection has confirmed
that all of them have signatures of emission in their line profiles.
Some have even prominent strong emissions.
These include three supernovae candidates, an unknown Wolf-Rayet star,
and many Be stars and young stellar objects.
Section~\ref{examples} shows examples of our findings.

Moreover, through the visual preview of candidates, several normal and Seyfert galaxies and a high-velocity star, LAMOST HVS1, were also identified.
Lastly, we compared our active deep learning method to the dual non-active learning scenario
in Appendix~\ref{appendix:non-active}.
The comparison shows the significant gain of our active deep learning method.

\subsection{On-line material}

The final catalogues of spectra of all our emission-line candidates obtained
by active deep learning are available at the CDS (see the footnote on the title page),
and they are also stored in the science cloud in the
\href{https://doi.org/10.5281/zenodo.3241520}{Zenodo
repository}\footnote{\url{https://doi.org/10.5281/zenodo.3241520}}~\citep{zenodotables} for further investigation.   HTML
tables list the spectra that were known by H16, spectra of new objects that we were
able to cross-match with SIMBAD, and all our new so far unknown
emission-line spectra.  For the sake of completeness, we also show the table
of spectra that were visually proved to be in some way broken, extremely noisy, or lacked emission
signatures.  All of these tables also contain direct links to the
CDS Vizier repository of LAMOST DR2, where the particular spectrum may be
interactively plotted.  Furthermore, there are the CSV versions (suitable for
spreadsheets) and VOTable format for further analysis in VO tools, such as
Aladin, Topcat, or SPLAT-VO.
The electronic PDF version of this paper includes electronic links to various public
resources, such as LAMOST or SDSS archives of spectra, DSS images
of the sky, or detailed description of objects in SIMBAD including links to
our previews stored at \href{https://doi.org/10.5281/zenodo.3236165}{Zenodo}\footnote{\url{https://doi.org/10.5281/zenodo.3236165}} as
well.

\subsection{Examples of rare interesting objects}
\label{examples}

The important research result is also a list of objects with unusual spectra properties.
Some objects may have been caught just during a LAMOST observation in a particular evolutionary phase,
or we might have witnessed the outburst of an unknown nova or supernova.
Many such spectra correspond to known SIMBAD objects,
but the SIMBAD class of such objects may be different from
what we see.
Very interesting objects were also found by checking the bad (noisy or artefact-contaminated) spectra in detail.
For example, a part of the spectral range could be destroyed,
but the remaining part with a dominant line profile may be still intact.

\subsubsection{New emission-line stars}

\begin{figure}
    \resizebox{\hsize}{!}{
        \includegraphics{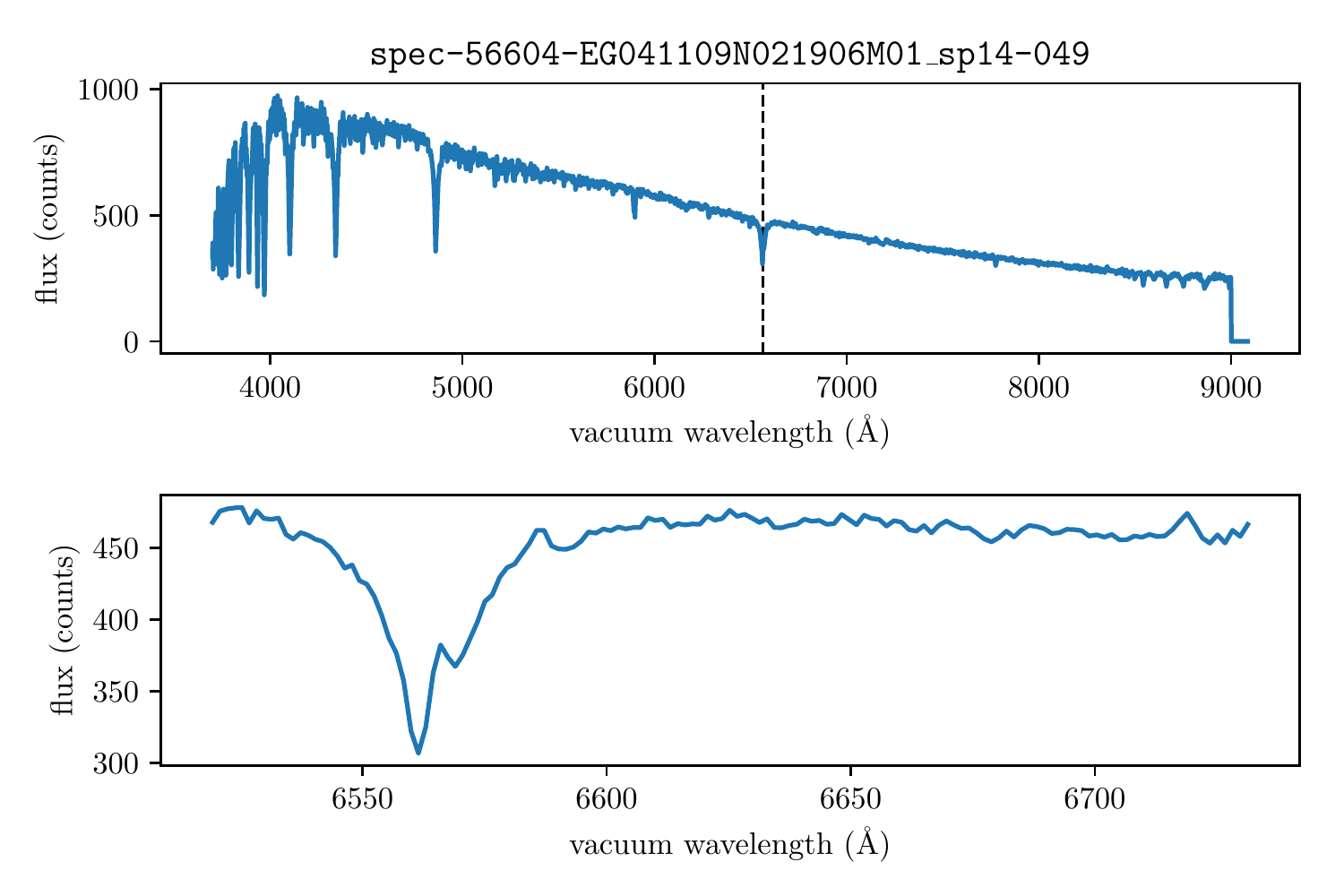}}
    \resizebox{\hsize}{!}{
        \includegraphics{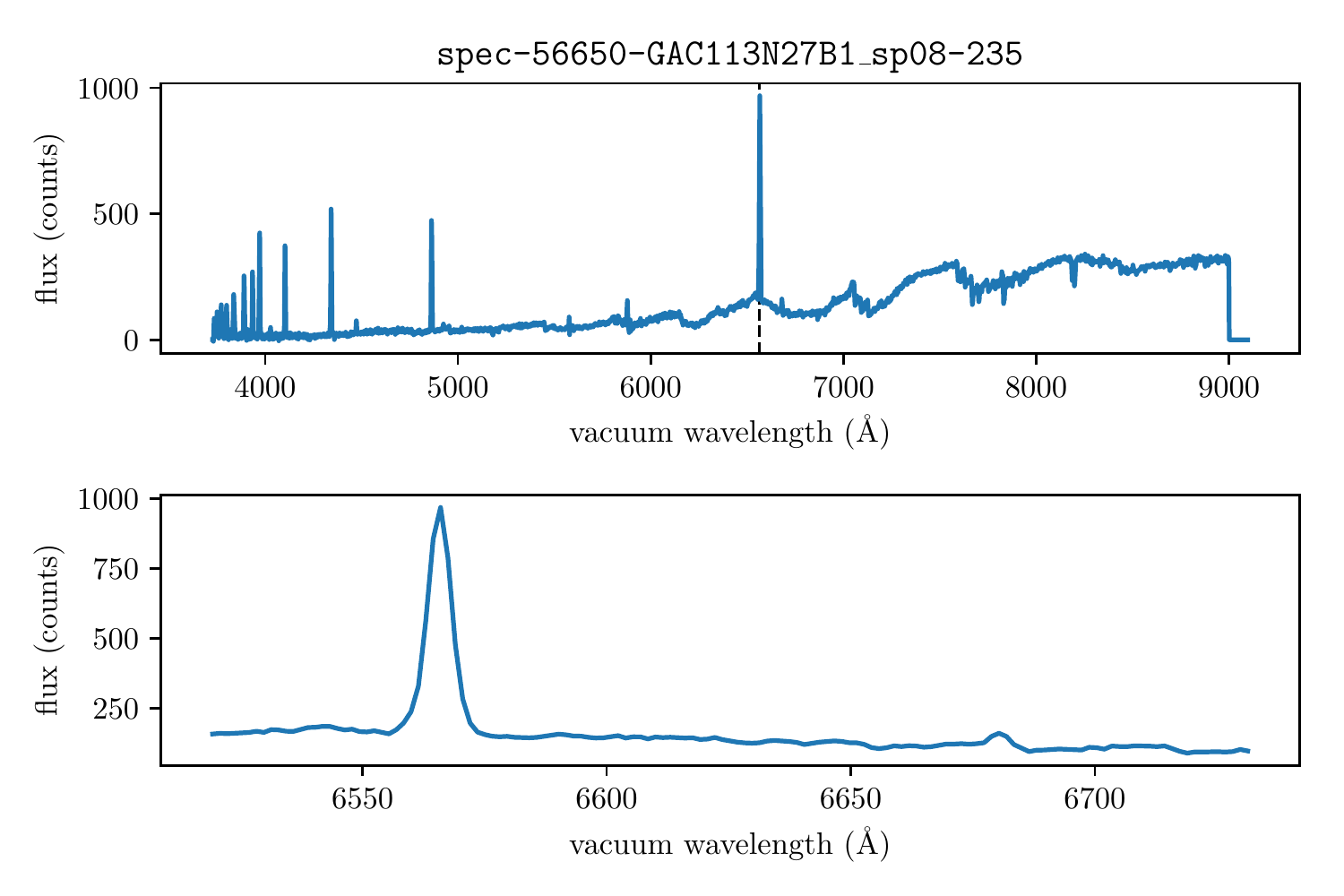}}
    \resizebox{\hsize}{!}{
        \includegraphics{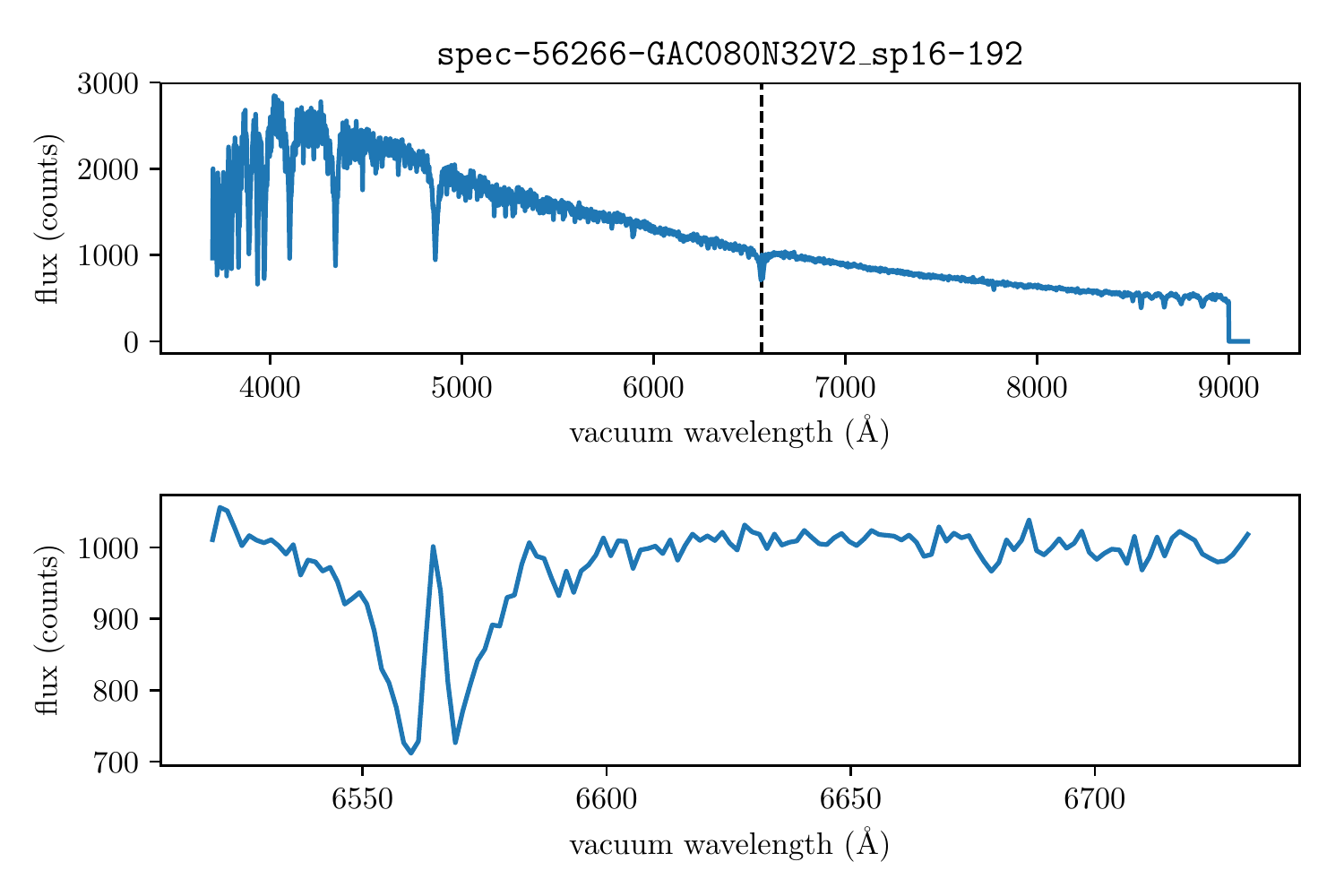}}
    \caption{Examples of spectra classified as single peaks.}
    \label{fig:examples_single}
\end{figure}

\begin{figure}
    \resizebox{\hsize}{!}{
        \includegraphics{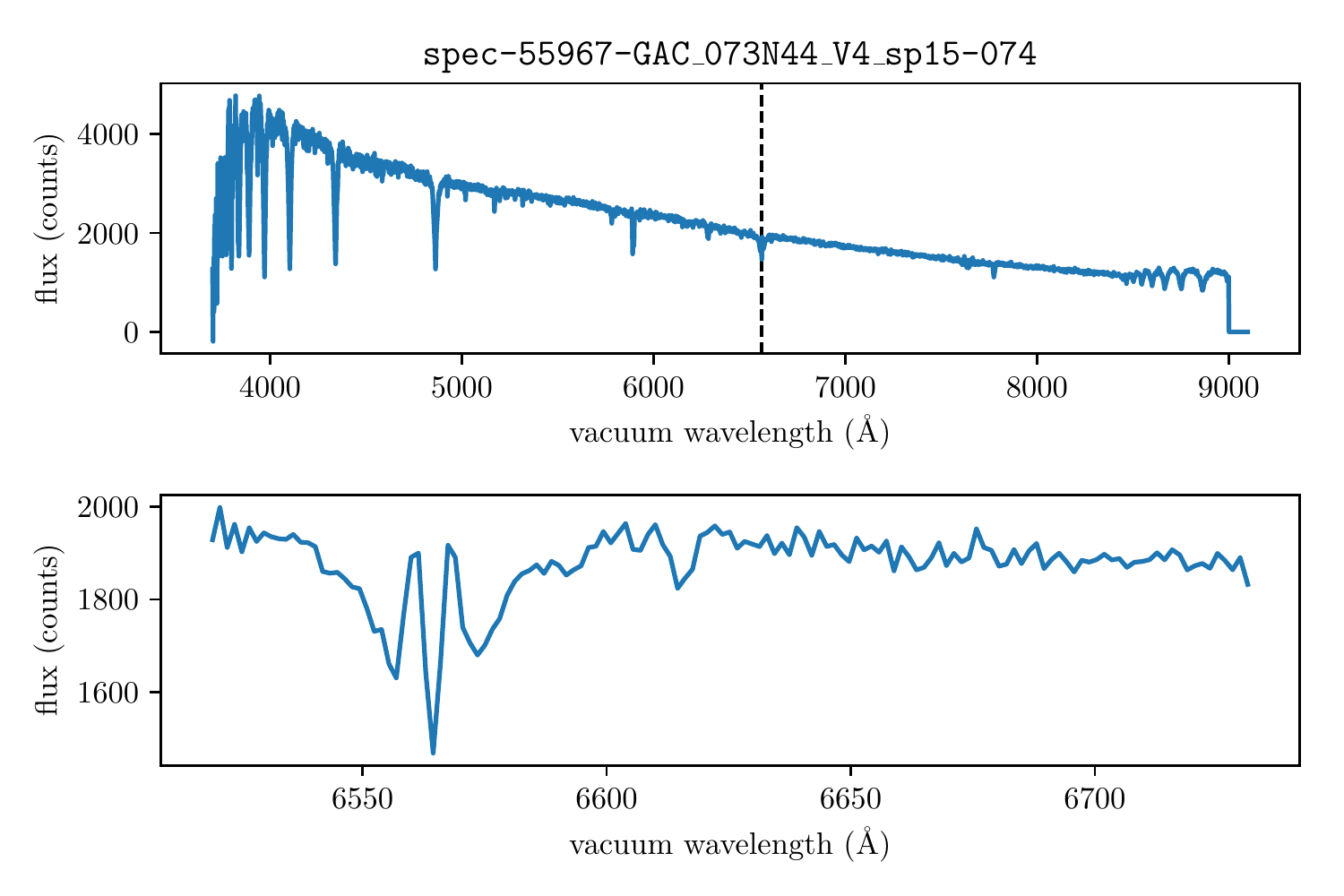}}
    \resizebox{\hsize}{!}{
        \includegraphics{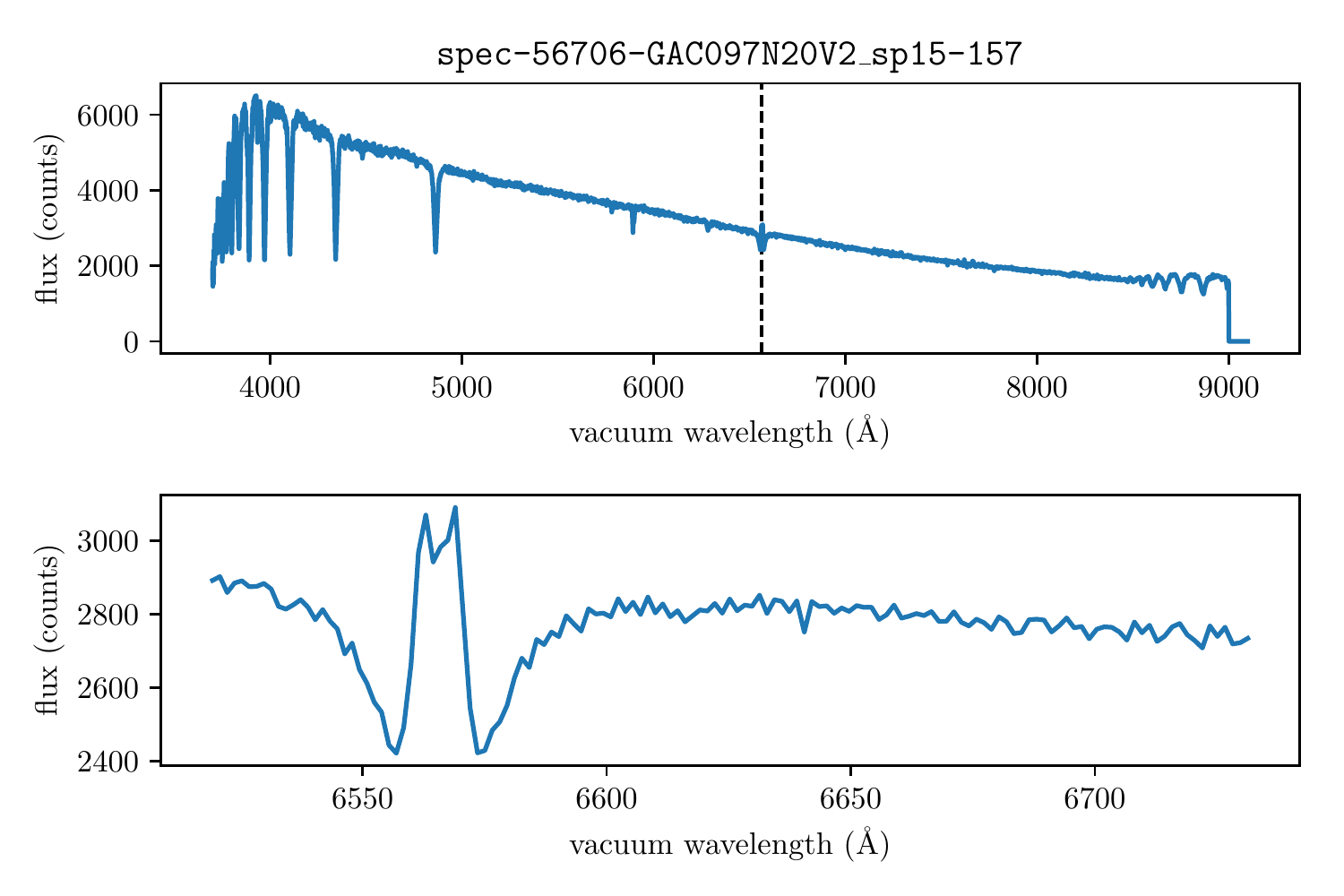}}
    \resizebox{\hsize}{!}{
        \includegraphics{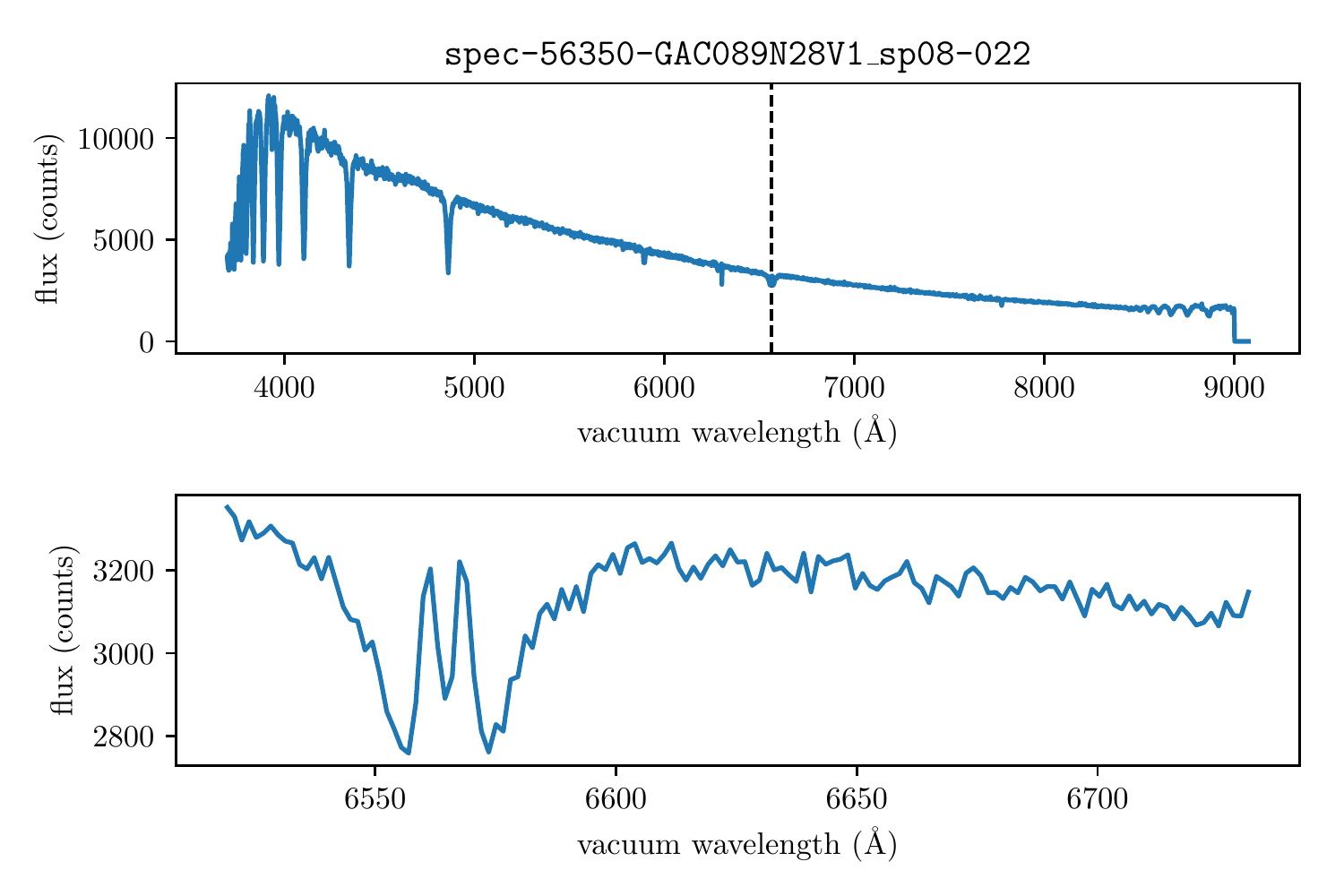}}
    \resizebox{\hsize}{!}{
        \includegraphics{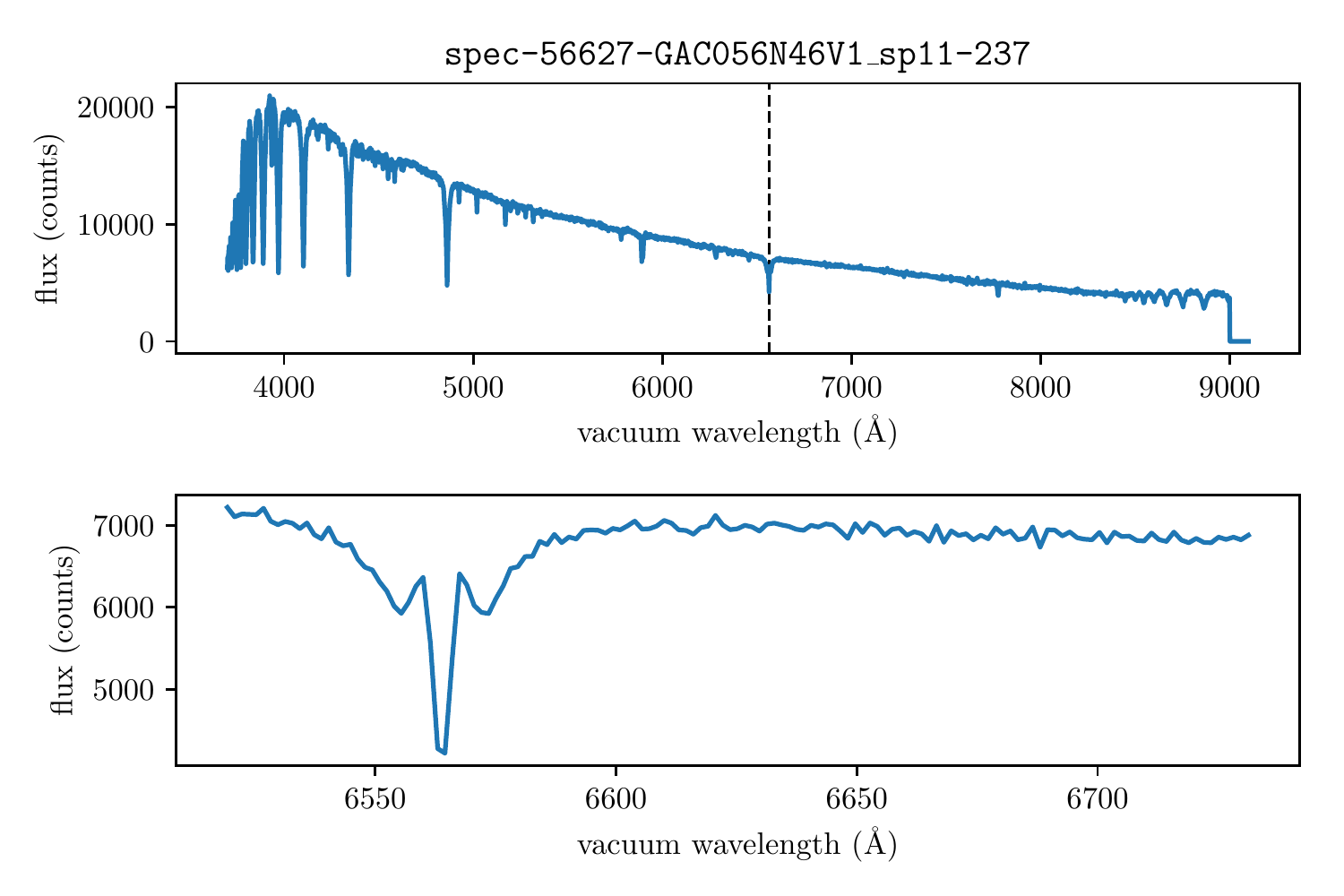}}
    \caption{Examples of spectra classified as double peaks.}
    \label{fig:examples_double}
\end{figure}

Our candidate list also includes 1\,013 spectra of 948 objects
that are neither cross-matched with SIMBAD nor listed by H16.
They cover a wide range of line profile shapes belonging to new so far unknown Be stars, T~Tau stars, cataclysmic variables, close binaries, symbiotic stars etc.
Figures~\ref{fig:examples_single}~and~\ref{fig:examples_double} show randomly selected examples of interesting spectra of each target class.

Some stars have quite peculiar spectra that exhibit quite
complicated profiles in multiple lines. This promises interesting physical
conditions.

\begin{figure}
    \resizebox{\hsize}{!}{
        \includegraphics{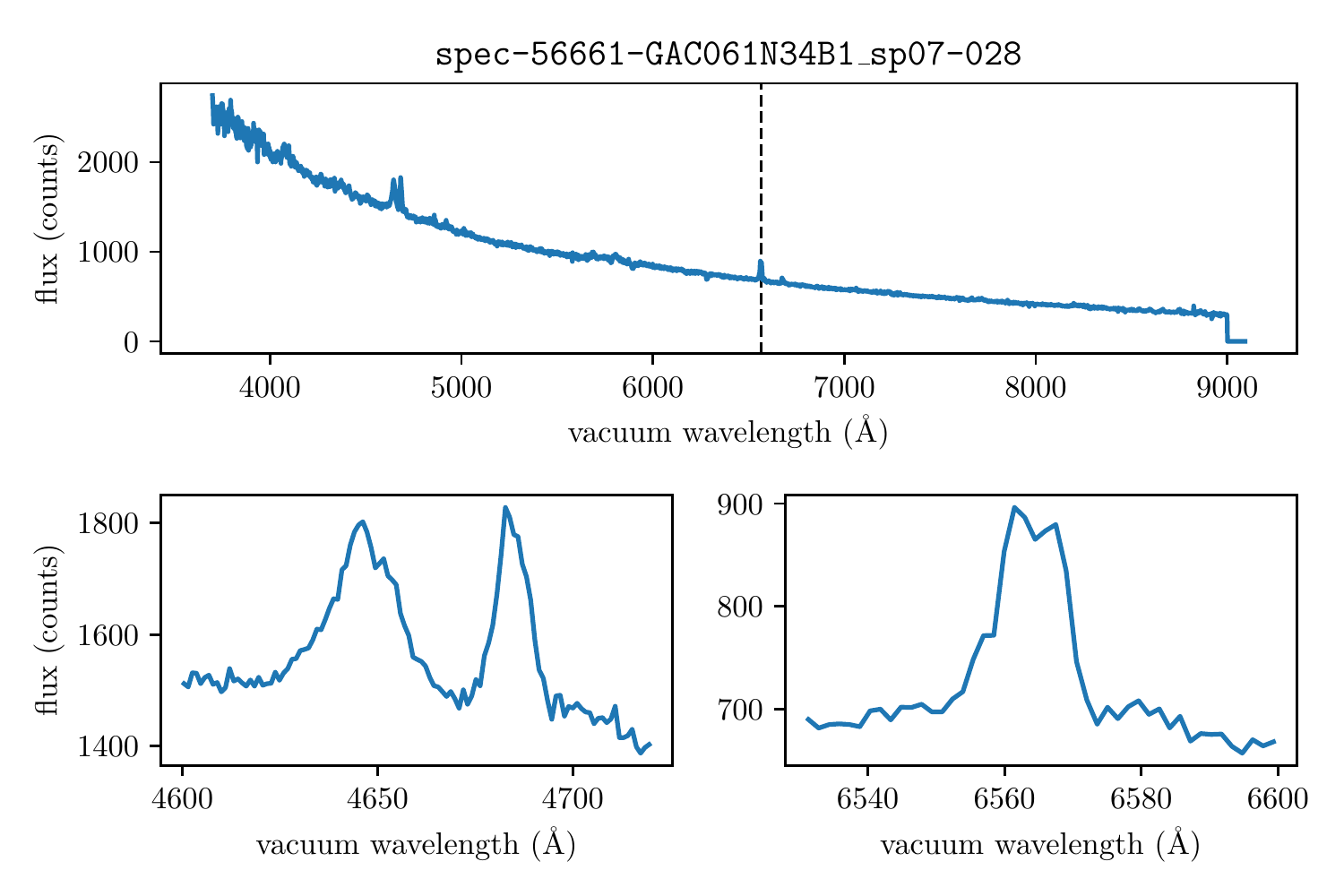}}
    \caption{Candidate Wolf-Rayet WN star LAMOST J040901.83+323955.6.}
    \label{WNstar}
\end{figure}

Spectrum
\href{https://zenodo.org/record/3236166/files/spec-56661-GAC061N34B1\_sp07-028.pdf}{\small\textsf{spec-56661-GAC061N34B1\_sp07-028}}
of object \href{http://dr2.lamost.org/spectrum/view?obsid=203507028}{LAMOST
J040901.83+323955.6} displays dominating emission in~\ion{N}{III}~4\,644~\AA{}
and \ion{He}{II}~4\,686~\AA{} lines, which suggests that this might be a
Wolf-Rayet WN star~\citep{Walborn_2000}.
See Fig.~\ref{WNstar}.

\begin{figure}
    \resizebox{\hsize}{!}{
        \includegraphics{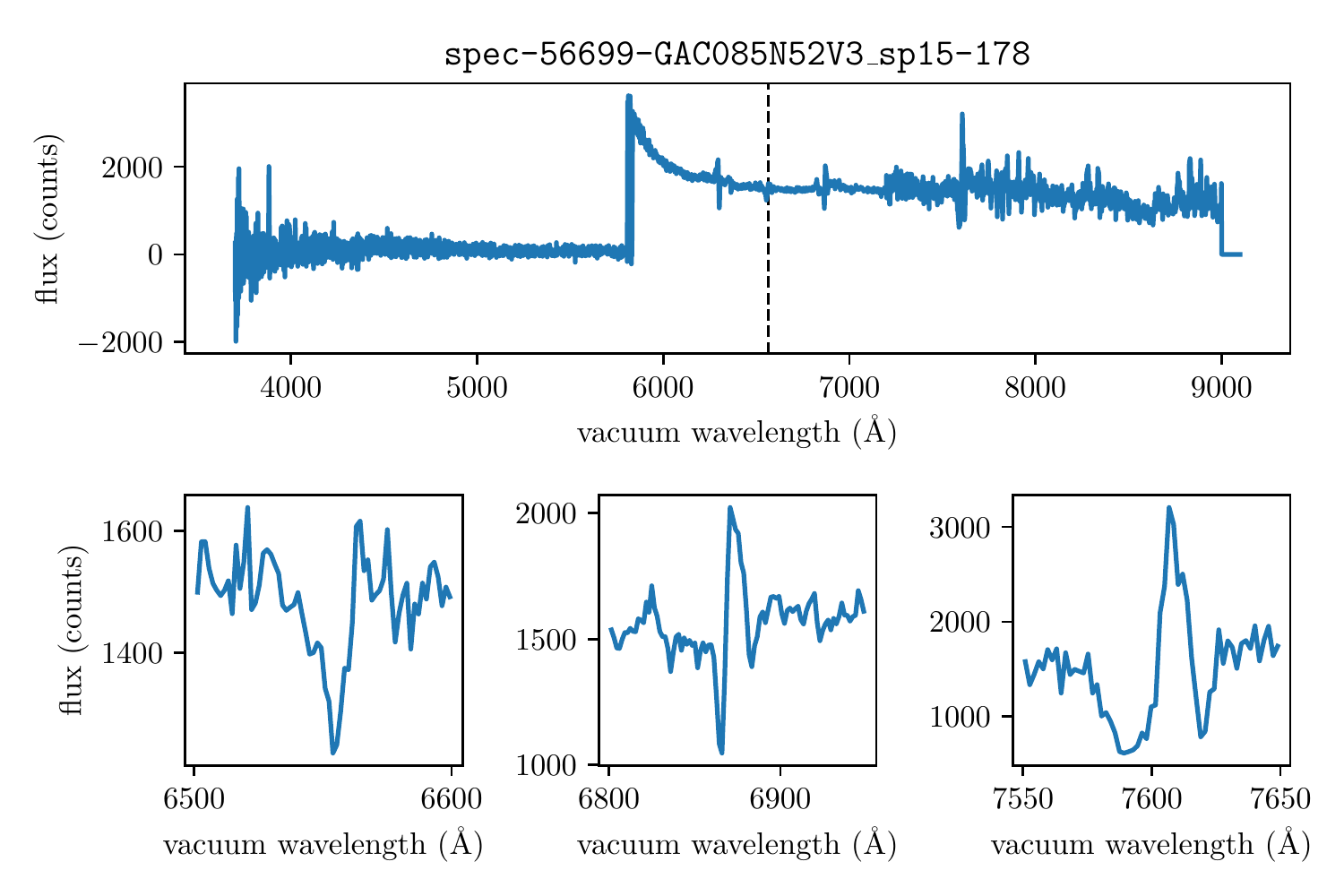}}
    \caption{Object LAMOST J053944.81+531825.7 with a probable gravitation lens.}
    \label{spec-56699-GAC085N52V3_sp15-178}
\end{figure}

Spectrum \href{https://zenodo.org/record/3236166/files/spec-56699-GAC085N52V3\_sp15-178.pdf}{\small\textsf{spec-56699-GAC085N52V3\_sp15-178}}
of object \href{http://dr2.lamost.org/spectrum/view?obsid=216415178}{LAMOST J053944.81+531825.7}
shows quite a bad merging of LAMOST red and blue spectrographs. This was done by the reduction pipeline
(therefore it is listed in the table of bad candidates),
but the red part shows P~Cyg and inverse P~Cyg profiles in several lines
and emission combined with absorption at 7\,610~\AA{}
(see~Fig.~\ref{spec-56699-GAC085N52V3_sp15-178}).
In the \href{http://archive.eso.org/dss/dss/image?ra=05+39+44.81&dec=53+18+25.7&equinox=J2000&name=&x=1&y=1&Sky-Survey=DSS2-red&mime-type=image%2Fgif&statsmode=WEBFORM}{DSS2}
survey, a strange object with the signature of an edge-on ring (resembling Saturn) lies at the given position.
However, the object is resolved into three in-line stars in~the~\href{https://ps1images.stsci.edu/cgi-bin/ps1cutouts?pos=05+39+44.81+%2B53+18+25.7}{PanSTARRS-1 survey}.
Both satellite stars are perfectly aligned in a straight line with the bright
central object, and they are separated by almost exactly 6.7\arcsec{} from its centre.
The configuration did not change since the PanSTARRS exposures (secured  \href{https://ps1images.stsci.edu/cgi-bin/ps1cutouts?pos=05+39+44.81+%2B53+18+25.7&&filetypes=warp}{multiple times}  in years 2011 to 2014) up to now, which we confirmed on 15 April 2020 by the \href{https://zenodo.org/record/3959230}{exposure} of the newly commissioned photometric camera of the 2m Perek telescope at Ond\v{r}ejov observatory.  
The symmetrical configuration deserves further investigation as it may be an effect of gravitational lensing.

\begin{figure}
    \resizebox{\hsize}{!}{
        \includegraphics{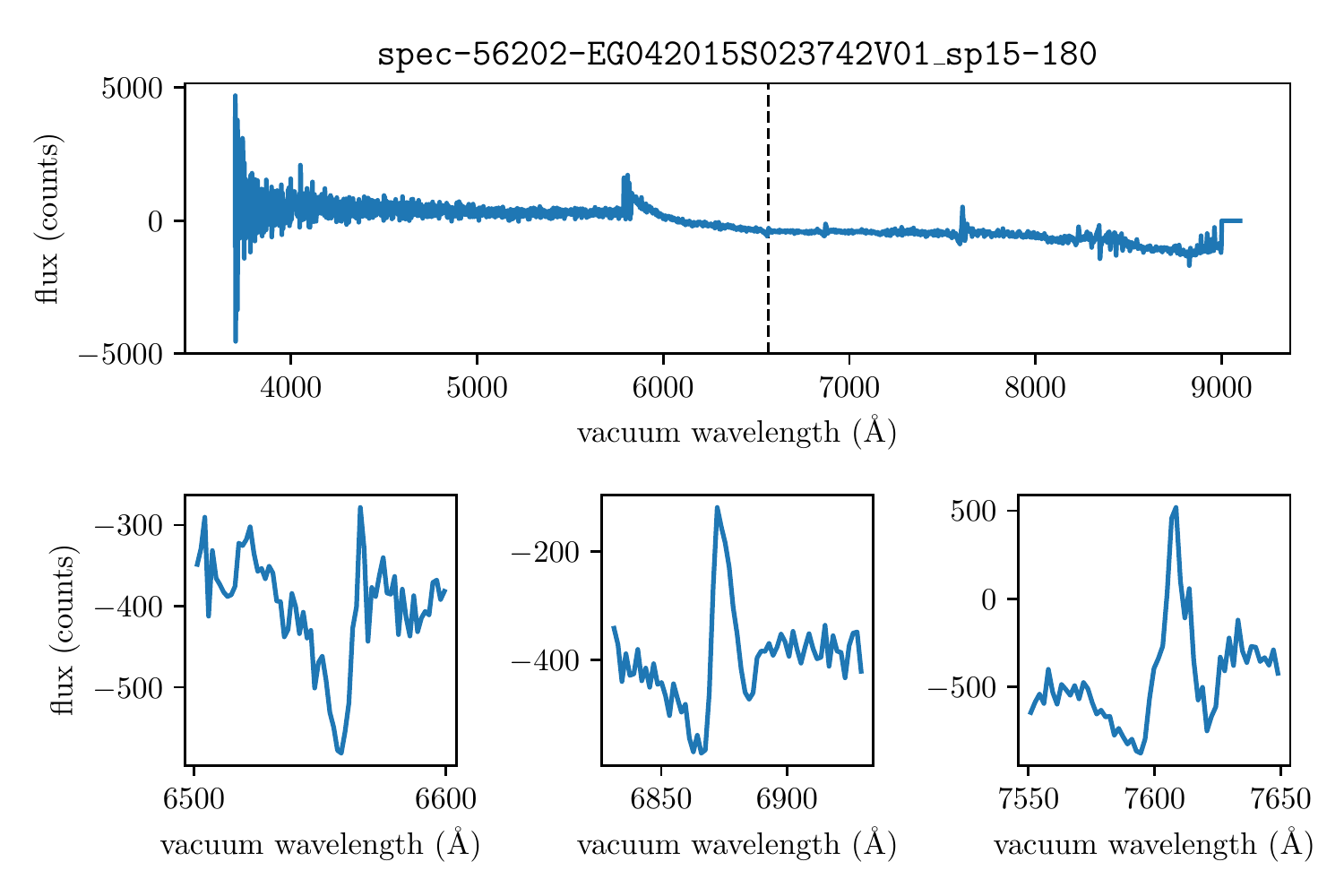}}
    \caption{Object LAMOST J041919.80-020211.6 with complex profiles.}
    \label{spec-56202-EG042015S023742V01_sp15-180}
\end{figure}

Spectrum \href{https://zenodo.org/record/3236166/files/spec-56202-EG042015S023742V01\_sp15-180.pdf}{\small\textsf{spec-56202-EG042015S023742V01\_sp15-180}}
in~Fig.~\ref{spec-56202-EG042015S023742V01_sp15-180} is very similar,
including the same bug in joining red and blue spectrograph,
showing the object \href{http://dr2.lamost.org/spectrum/view?obsid=57215180}{LAMOST J041919.80-020211.6},
which is neither known to SIMBAD nor exposed by the SDSS.
It has two emissions at 5\,790~\AA{} and 5\,820~\AA{},
the H\(\alpha\) absorption with an emission peak in the red wing,
and a strong emission combined with absorption at 7\,600~\AA{}.
A P~Cyg profile is visible at 6\,870~\AA{}.

\begin{figure}
    \resizebox{\hsize}{!}{
        \includegraphics{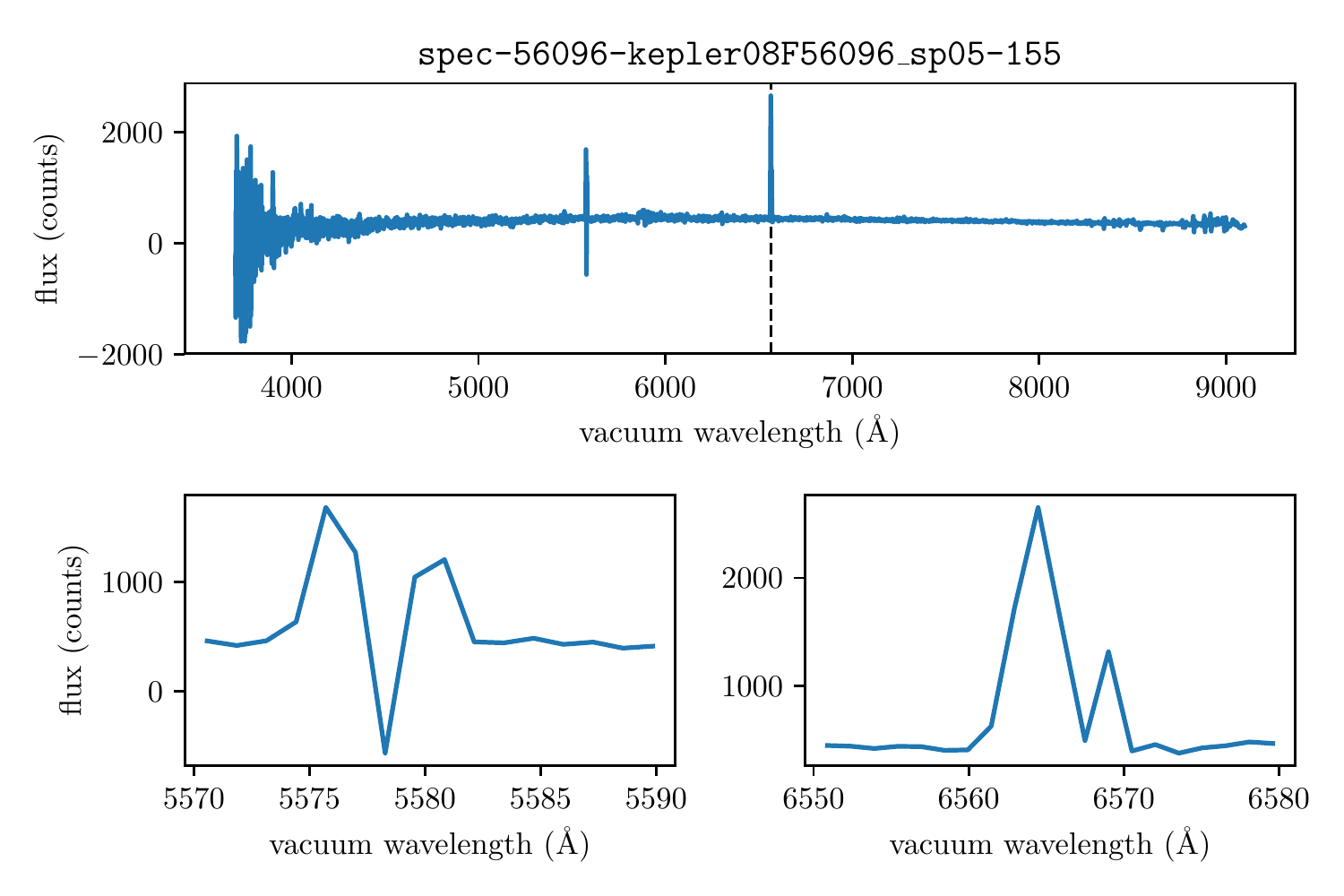}}
    \resizebox{\hsize}{!}{
        \includegraphics{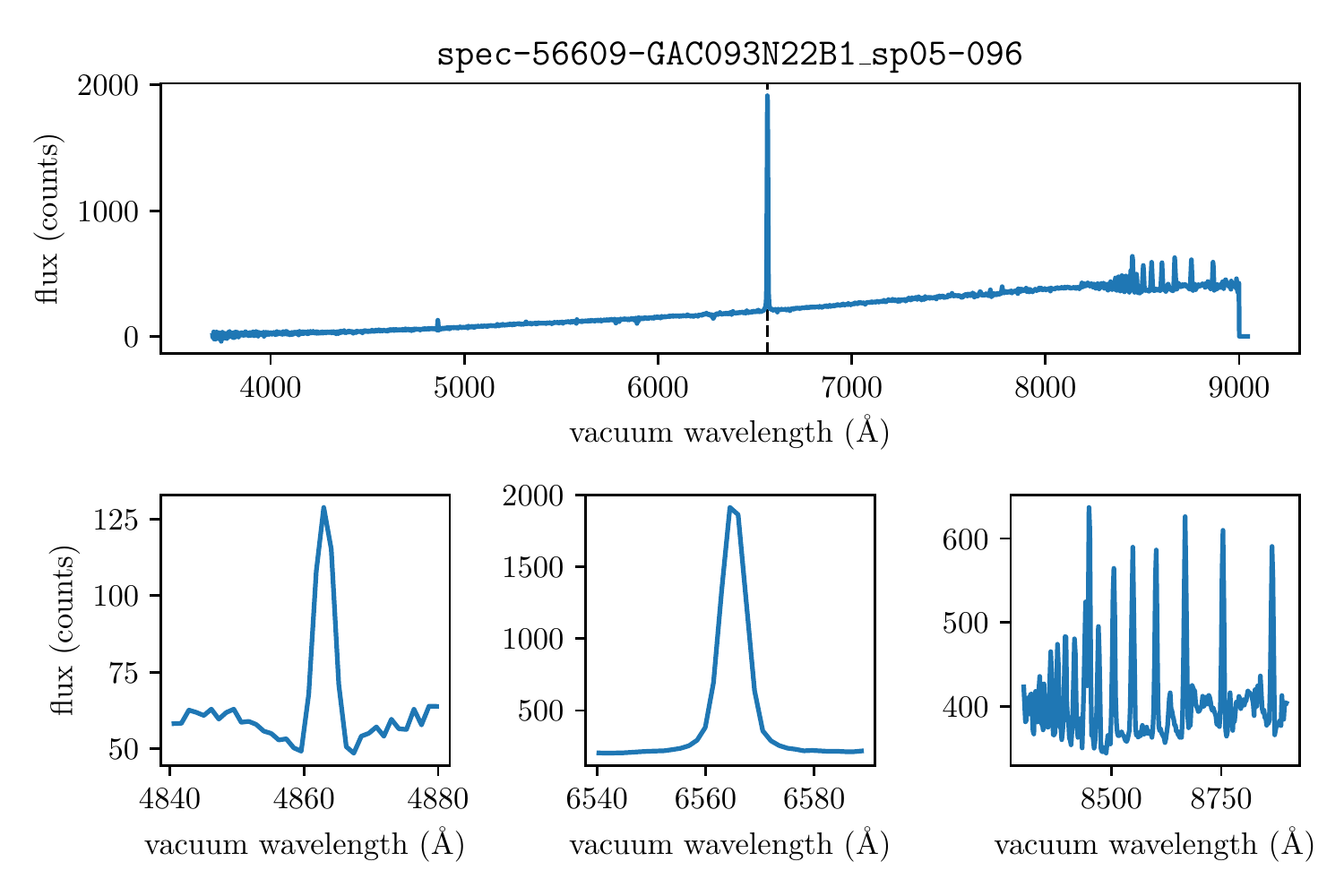}}
    \resizebox{\hsize}{!}{
        \includegraphics{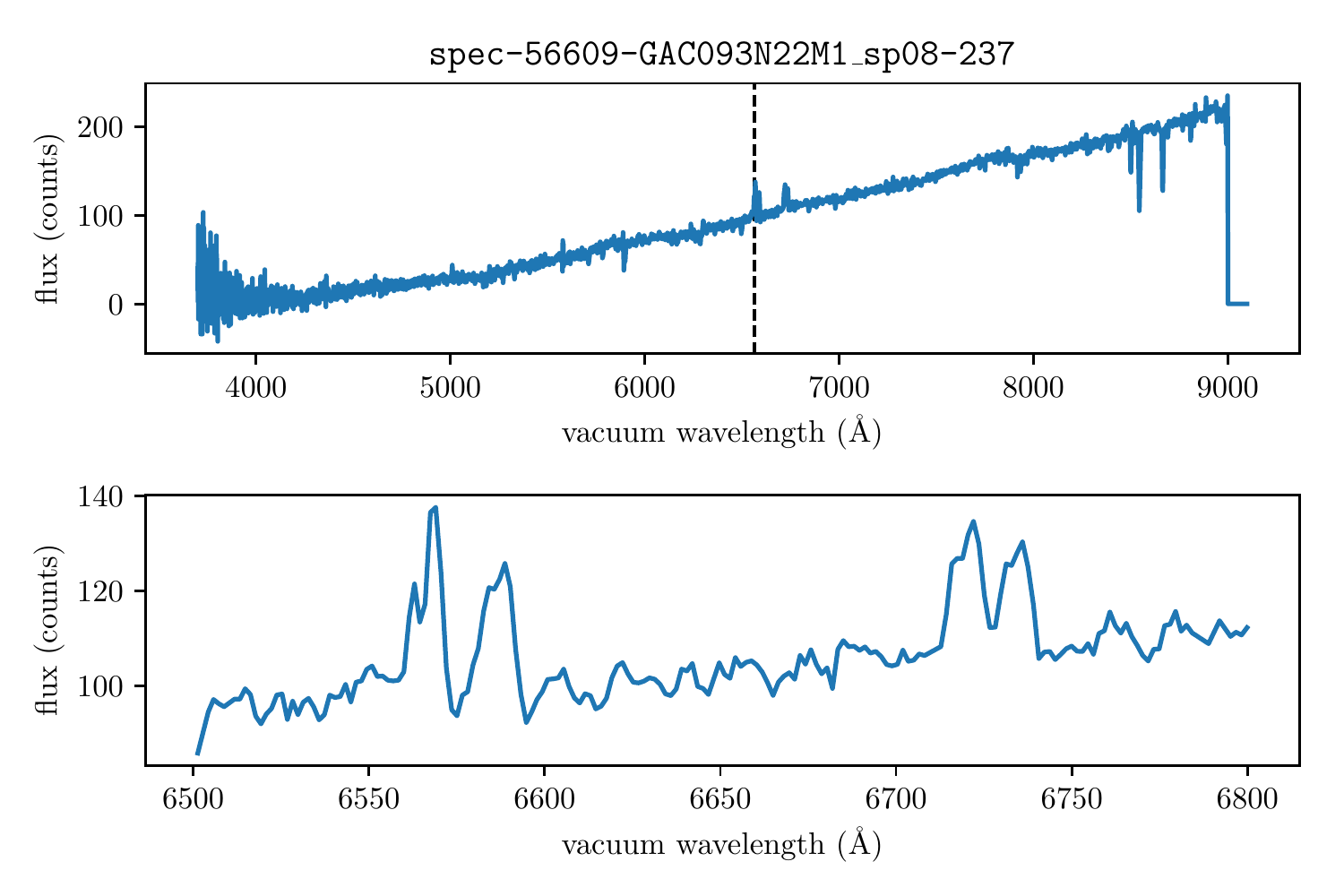}}
    \resizebox{\hsize}{!}{
        \includegraphics{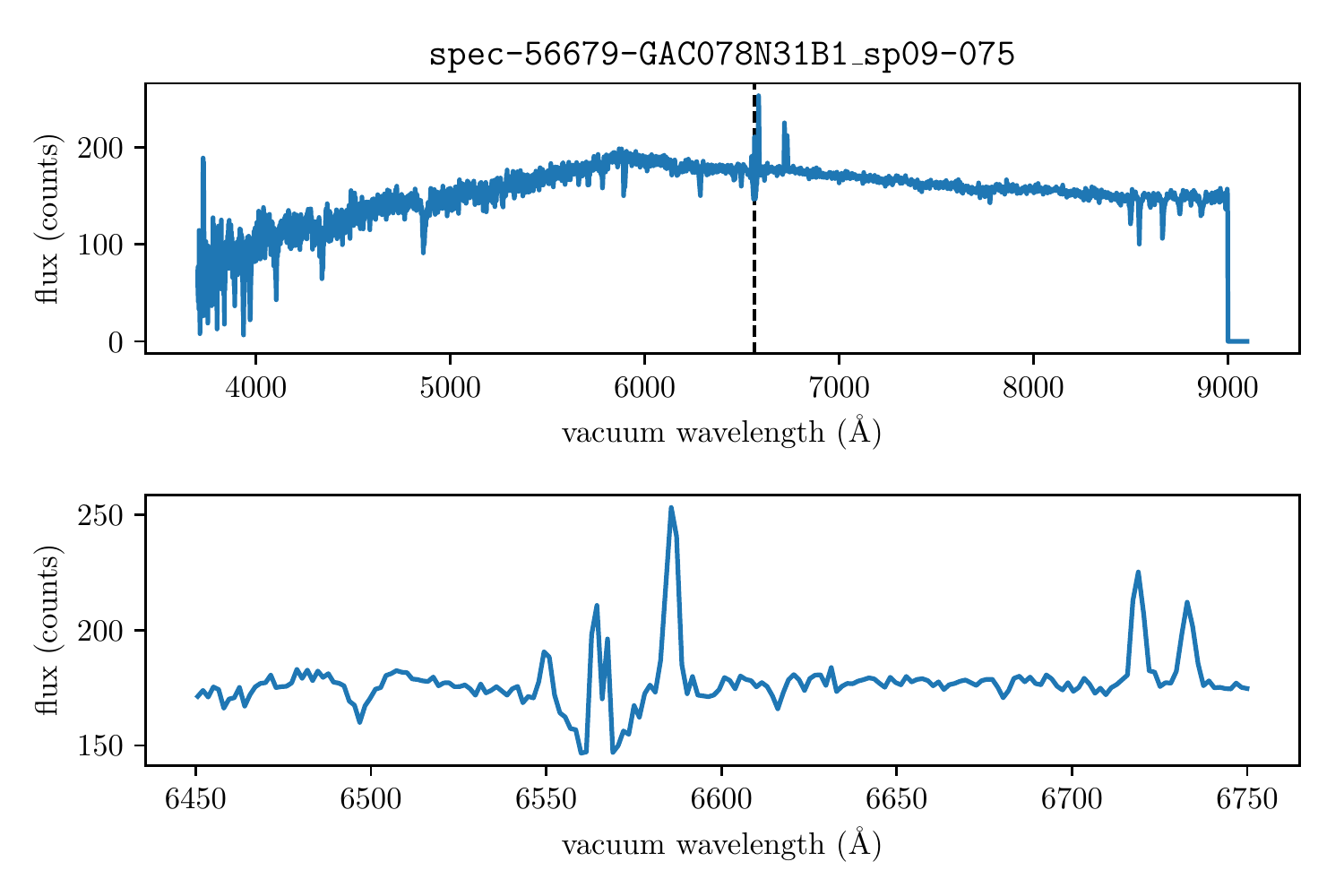}}
    \caption{Random discovered objects with complex profiles.}
    \label{complex_profiles}
\end{figure}

Figure~\ref{complex_profiles} gives examples of other newly discovered objects with complex spectra.

\subsubsection{Supernovae candidates}
\label{supernovae}

Three spectra with a very bright and wide single-peak emission
line with an FWHM of about 200~\AA\ are very unusual{}.
H16 identified none of these objects,
and they cannot be cross-matched with SIMBAD.
We have found all of them in SDSS DR15,
but no estimate of the spectral class is given there.
They are claimed to be stars, however.
Because of their extremely wide red-shifted H\(\alpha\) line
(the others are hidden in noise)
and because several galaxies are seen around them,
we speculate that they may be distant supernovae.
However, they might be mere reduction artefacts as well,
therefore further investigation is desirable.

Spectrum \href{https://zenodo.org/record/3236166/files/spec-56012-F5601204\_sp10-030.pdf}{\small\textsf{spec-56012-F5601204\_sp10-030}}
belongs to object \href{http://dr2.lamost.org/spectrum/view?obsid=48110030}{LAMOST J114232.73-011535.9}.
In the SDSS DR15 colour image at this position, a faint white star is visible that may be identified with
\href{http://skyserver.sdss.org/DR15/en/tools/explore/summary.aspx?ra=175.636393&dec=-1.259991}{SDSS J114232.73-011535.9},
\href{http://skyserver.sdss.org/DR15/en/tools/explore/neighbors.aspx?id=1237671128588222648&spec=&apid=&fieldId=0x112d175580420000&ra=175.63639393005&dec=-1.259991555891}{surrounded} by galaxies.

Spectrum \href{https://zenodo.org/record/3236166/files/spec-56012-F5601204\_sp02-158.pdf}{\small\textsf{spec-56012-F5601204\_sp02-158}}
is the \href{http://dr2.lamost.org/spectrum/view?obsid=48102158}{LAMOST J114009.42-012454.3} equivalent to
\href{http://skyserver.sdss.org/DR15/en/tools/explore/summary.aspx?ra=175.039289&dec=-1.415085}{SDSS J114009.42-012454.3}.
A faint orange object is
\href{http://skyserver.sdss.org/DR15/en/tools/explore/neighbors.aspx?id=1237671129124897166&spec=&apid=&fieldId=0x112d1755a03f0000&ra=175.039288127368&dec=-1.41508524671862}{surrounded}
by a number of galaxies seen within 10\arcsec{} in the SDSS.

The last spectrum of a similar shape is
\href{https://zenodo.org/record/3236166/files/spec-56396-HD165712N321400M01\_sp06-166.pdf}{\small\textsf{spec-56396-HD165712N321400M01\_sp06-166}}
of object \href{http://dr2.lamost.org/spectrum/view?obsid=144706166}{LAMOST J170758.50+313441.1}.
Its counterpart in the SDSS is seen as a blue circular object,
\href{http://skyserver.sdss.org/DR15/en/tools/explore/summary.aspx?ra=256.993779&dec=31.578089}{SDSS
J170758.50+313441.1,}
again with the nearby galaxy
\href{http://skyserver.sdss.org/DR15/en/tools/explore/Summary.aspx?id=1237655473973166122}{SDSS
J170758.05+313443.6} about 6\arcsec{} apart.
The object is also present in the 2MASS and GALEX surveys.

\begin{figure}
    \resizebox{\hsize}{!}{
        \includegraphics{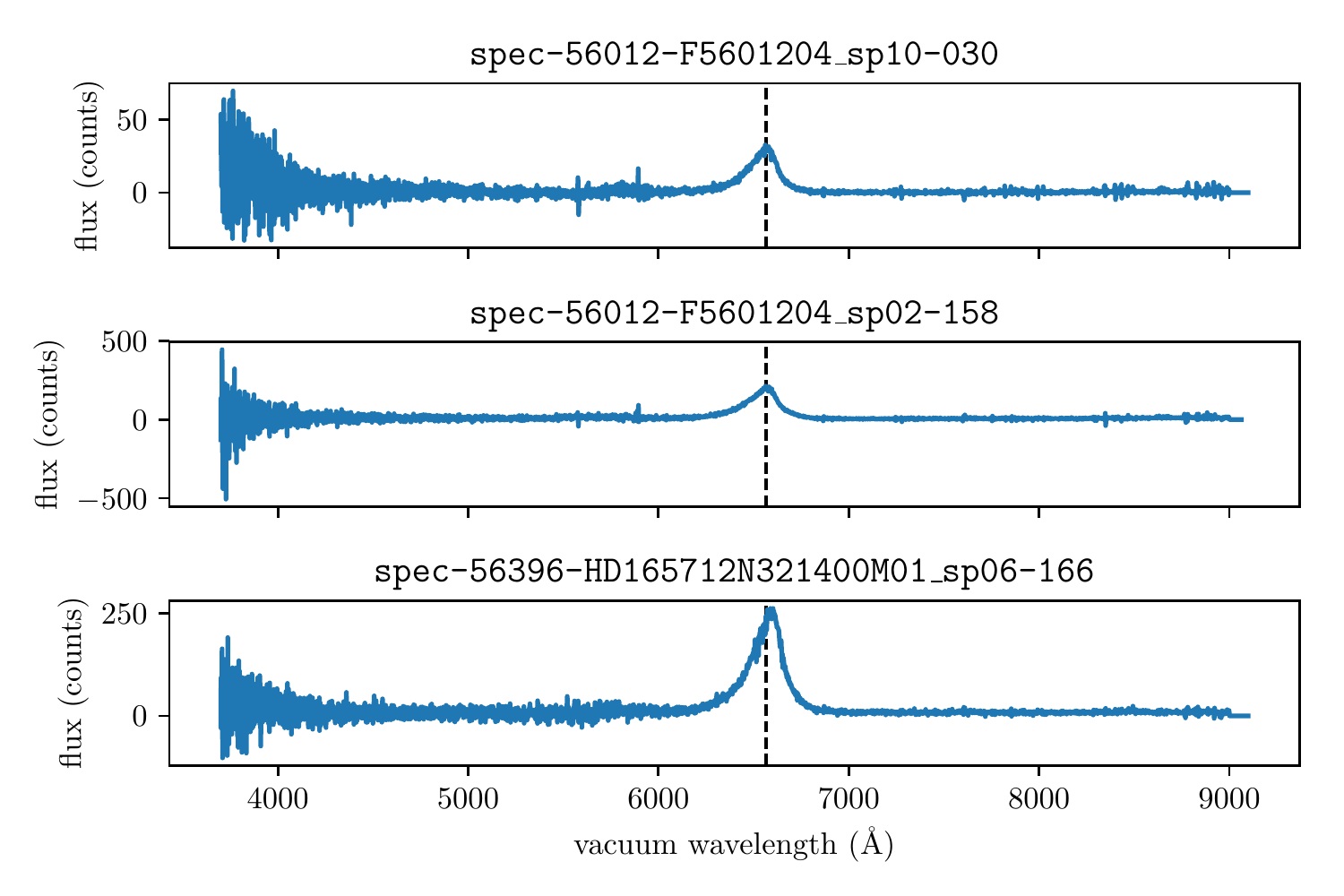}}
    \caption{Supernovae candidates identified in the LAMOST DR2.}
    \label{supernova_candidates}
\end{figure}

\subsubsection{Detection of extragalactic objects}

Visual inspection of suggested candidates also resulted in the identification of several extragalactic objects,
both Seyfert and normal galaxies (confirmed by SIMBAD).

Spectrum \href{https://zenodo.org/record/3236166/files/spec-56716-HD114322N280318M\_sp08-042.pdf}{\small\textsf{spec-56716-HD114322N280318M\_sp08-042}}
belongs to object \href{http://dr2.lamost.org/spectrum/view?obsid=218908042}{LAMOST J114631.67+274624.1},
which is the Seyfert galaxy \href{http://simbad.u-strasbg.fr/simbad/sim-coo?Coord=176.63197+27.77338&CooFrame=FK5&CooEpoch=2000&CooEqui=2000&CooDefinedFrames=none&Radius=10&Radius.unit=arcsec}{87GB 114356.6+280254} with redshift 0.3187.

Spectrum \href{https://zenodo.org/record/3236166/files/spec-56657-M31020N36M1\_sp08-216.pdf}{\small\textsf{spec-56657-M31020N36M1\_sp08-216}}
belongs to object \href{http://dr2.lamost.org/spectrum/view?obsid=201408216}{LAMOST
J012555.94+351036.9},
which is the Seyfert galaxy \href{http://simbad.u-strasbg.fr/simbad/sim-coo?Coord=21.483087+35.17692&Radius=2&Radius.unit=arcsec}{2MASS
J01255593+3510368,} for which no SDSS spectrum is available.

Spectrum \href{https://zenodo.org/record/3236166/files/spec-56798-HD141746N331518M01\_sp16-150.pdf}{\small\textsf{spec-56798-HD141746N331518M01\_sp16-150}}
belongs to object \href{http://dr2.lamost.org/spectrum/view?obsid=240716150}{LAMOST
J141403.15+352311.3}
which is the Seyfert galaxy \href{http://simbad.u-strasbg.fr/simbad/sim-coo?Coord=213.51314++35.386499&CooFrame=FK5&CooEpoch=2000&CooEqui=2000&CooDefinedFrames=none&Radius=2&Radius.unit=arcsec}{2MASX J14140315+3523107}.
The \href{http://skyserver.sdss.org/DR15//en/get/SpecById.ashx?id=1849887134777894912}{SDSS spectrum} is also available.

Spectrum \href{https://zenodo.org/record/3236166/files/spec-56752-HD150254N020528B01\_sp08-150.pdf}{\small\textsf{spec-56752-HD150254N020528B01\_sp08-150}}
belongs to object \href{http://dr2.lamost.org/spectrum/view?obsid=232608150}{LAMOST J150711.68+013202.4},
which is a star in the outer part of the LINER-type galaxy
\href{http://simbad.u-strasbg.fr/simbad/sim-coo?Coord=15+07+11.68224000+\%2B01+32+02.4720000&Radius=2&Radius.unit=arcmin}{NGC5850}.

Spectrum \href{https://zenodo.org/record/3236166/files/spec-56304-HD083110N401329F01\_sp09-064.pdf}{\small\textsf{spec-56304-HD083110N401329F01\_sp09-064}}
belongs to object \href{http://dr2.lamost.org/spectrum/view?obsid=105909064}{LAMOST
J083426.80+411414.8}, which is not in SIMBAD, but it is cross-matched with SDSS DR15 as galaxy
\href{http://skyserver.sdss.org/DR15/en/tools/explore/summary.aspx?ra=128.6116844&dec=41.237464}{SDSS
J083426.80+411414.8}.
An \href{http://skyserver.sdss.org/DR12//en/get/SpecById.ashx?id=1007812401321502720}{SDSS spectrum} is also available.

Spectrum \href{https://zenodo.org/record/3236166/files/spec-56633-HD095000N333605M01\_sp10-109.pdf}{\small\textsf{spec-56633-HD095000N333605M01\_sp10-109}}
of object \href{http://dr2.lamost.org/spectrum/view?obsid=188210109}{LAMOST
J093915.39+331634.2}
belongs to galaxy \href{http://simbad.u-strasbg.fr/simbad/sim-coo?Coord=144.81416+33.27617&Radius=5&Radius.unit=arcsec}{LEDA~2028208}
with an available \href{http://skyserver.sdss.org/DR15/en/get/SpecById.ashx?id=6535876003557777408}{SDSS spectrum}.

Spectrum \href{https://zenodo.org/record/3236166/files/spec-56739-HD114047N212109B\_sp10-131.pdf}{\small\textsf{spec-56739-HD114047N212109B\_sp10-131}}
of object \href{http://dr2.lamost.org/spectrum/view?obsid=228210131}{LAMOST
J113114.00+211841.3}
belongs to the galaxy \href{http://simbad.u-strasbg.fr/simbad/sim-coo?Coord=172.8083401+21.3114819&Radius=5&Radius.unit=arcsec}{LEDA
1647402} with an available \href{http://skyserver.sdss.org/DR15/en/get/SpecById.ashx?id=2819326554369189888}{SDSS spectrum}.

\subsubsection{High-velocity stars}
\label{HVS}

Identified as target class objects by our CNN
(despite the relatively minor deformation of the absorption line profile, which is comparable with a noise level), a group of objects has a considerable red- or blue-shifted H\(\alpha\) line.
It appears that most of them are stars labelled in the LAMOST observing program as M31 targets
coming from the observing plan that covered the central regions of the M31 and M33 galaxies~\citep{DR1}.
The spectra clearly show that the centre of the H\(\alpha\) emission line is heavily blue-shifted,
which corresponds to line-of-sight velocities of \(-301\)~km~s\(^{-1}\) and \(-180\)~km~s\(^{-1}\) of M31 and M33, respectively~\citep{van_der_Marel_2008}.

However, other objects are not associated with galaxies by the LAMOST observing plan.
The most red-shifted is the spectrum (available in our previews)
\href{https://zenodo.org/record/3236166/files/spec-56357-HD090901N073047B01\_sp12-133.pdf}{\small\textsf{spec-56357-HD090901N073047B01\_sp12-133}}
of object LAMOST J091206.52+091621.8,
which according to the SIMBAD position belongs to the galaxy run-away hypervelocity star \href{http://simbad.u-strasbg.fr/simbad/sim-coo?Coord=09+12+06.52000000+\%2B09+16+21.8000000&Radius=2&Radius.unit=arcsec}{LAMOST HVS1}
with a stated radial velocity of 611.65~km~s\(^{-1}\)~\citep{HVS1}.
This spectrum is classified in the LAMOST \href{http://dr2.lamost.org/spectrum/view?obsid=131012133}{DR2 archive}
as a galaxy, while in its FITS header,
where it comes from our preview above, as an A0III star.
Another spectrum
(\href{http://dr2.lamost.org/spectrum/view?obsid=93907117}{spec-56285-HD090744N104005B03\_sp07-117})
of the same object, classified in the DR2 archive as an A0III star, is available.

During the visual preview, we found another two objects with high red-shifted radial velocity.
Spectrum \href{https://zenodo.org/record/3236166/files/spec-55997-B5599703\_sp14-056.pdf}{\small\textsf{spec-55997-B5599703\_sp14-056}}
of object \href{http://dr2.lamost.org/spectrum/view?obsid=45514056}{LAMOST J105350.26+271352.2},
which cannot be identified with SIMBAD,
shows RV 331~km~s\(^{-1}\) in H\(\alpha\) and 293~km~s\(^{-1}\) in H\(\beta\).
Four other spectra of the same star are listed in DR2,
namely
\href{http://dr2.lamost.org/spectrum/view?obsid=15414056}{spec-55910-B91005\_sp14-056},
\href{http://dr2.lamost.org/spectrum/view?obsid=45814056}{spec-55998-B5599803\_sp14-056},
\href{http://dr2.lamost.org/spectrum/view?obsid=189308015}{spec-56638-HD104953N275826B01\_sp08-015,}
and \href{http://dr2.lamost.org/spectrum/view?obsid=212708015}{spec-56685-HD104953N275826M01\_sp08-015}.

Spectrum \href{https://zenodo.org/record/3236166/files/spec-56617-VB081S05V1\_sp02-071.pdf}{\small\textsf{spec-56617-VB081S05V1\_sp02-071}}
of object \href{http://dr2.lamost.org/spectrum/view?obsid=181102071}{LAMOST
J052354.52-070508.3}
gives radial velocities 358~km~s\(^{-1}\) in H\(\alpha\) and 351~km~s\(^{-1}\) in H\(\beta\).
It is the same in another spectrum, \href{http://dr2.lamost.org/spectrum/view?obsid=181202071}{spec-56617-VB081S05V2\_sp02-071}.

Spectrum \href{https://zenodo.org/record/3236166/files/spec-56393-HD172143N395828M01\_sp16-162.pdf}{\small\textsf{spec-56393-HD172143N395828M01\_sp16-162}}
of object \href{http://dr2.lamost.org/spectrum/view?obsid=143116162}{LAMOST J171623.21+412303.1} is quite surprising.
The H\(\alpha\) seems to be blue-shifted by about 710~km~s\(^{-1}\)
and the Paschen P13 and P15 lines by 660~km~s\(^{-1}\) and 630~km~s\(^{-1}\), respectively.
Because of the high noise and apparent asymmetry of the lines,
it is impossible to obtain a precision better than about 10~km~s\(^{-1}\),
but it is evident that this star is one of the fastest stars in our galaxy and
approaches us, unlike HVS1.
However, it may be of extragalactic origin as well,
as the image of the object in \href{http://skyserver.sdss.org/DR15//en/tools/explore/summary.aspx?id=1237665583795404952}{SDSS DR15}
is partly saturated
and lies less than 5\arcsec{} away from the large elliptical galaxy.
It is known in the GALEX survey as GALEXASC J171623.29+412304.2, and in 2MASS as 2MASS J17162320+4123031.



\section{Conclusions}
\label{conclusion}

We have introduced a promising method for the discovery of objects of interest in large archives based on active deep learning.
This technique, supported by interactive visual classification of a small sample of suggested target classes, is very efficient and has led to the discovery of many new unknown emission-line stars.

To the best of our knowledge,
this is the first application of active deep learning techniques in
astronomy: used for spectral classification.
Many details still need to be elaborated,
and more experiments must be run on different samples of various types of spectra.
The main advantage of the method is that target classes with characteristic spectral features can be identified in cases
where the classical deep learning fails because not enough labelled examples are available.

Our experiments identified many emission-line candidates
that deserve more detailed examination because they may hide rare astronomical objects with interesting physical
properties.
All results are publicly available at Zenodo
and will also be uploaded in the main astronomical catalogue repository, the CDS Vizier.


\begin{acknowledgements}
    The early stages of this research were supported by the grant LD-15113 of
    Ministry of Education, Youth, and Sports of the Czech Republic and the
    COST Action TD1403 Big Sky Earth. The final parts of the research and
    all the experiments described here were supported by the same Ministry in project
    Research Center for Informatics, CZ.02.1.01/0.0/0.0/16\_019/0000765
    and by the Grant Agency of the Czech Technical University in Prague, grant No. SGS20/212/OHK3/3T/18.

    Furthermore, this work is based on spectra from the Ondřejov 2~m Perek
    telescope and the public LAMOST DR2 survey.
    Therefore, we would like to thank the Guoshoujing Telescope.
    Guoshoujing Telescope (the Large Sky Area
    Multi-Object Fiber Spectroscopic Telescope LAMOST) is a National Major
    Scientific Project built by the Chinese Academy of Sciences. Funding
    for the project has been provided by the National Development and
    Reform Commission. LAMOST is operated and managed by the National
    Astronomical Observatories, Chinese Academy of Sciences.

    We are namely indebted to Dr.~Chenzhou Cui for long-term support
    of our activities in the framework of collaboration with Chinese VO.

    Finally, we also express our thanks to Miroslav Šlechta for reduction
    of all CCD700 spectra used in our research.
\end{acknowledgements}

\bibliographystyle{aa}
\bibliography{references}


\begin{appendix}

\section{Comparison with non-active learning}
\label{appendix:non-active}

To clarify the real gain of active learning,
we compare our active deep learning method to a non-active learning dual scenario.
The non-active learning can be considered the zeroth iteration of the application of our active deep learning in~Sect.~\ref{experiments}.
However, the zeroth iteration in our application is carried out in the speeded-up regime
(see the last paragraph of Sect.~\ref{application}).

We carried out an independent experiment to prove the benefits of active learning.
We trained our CNN using the setting of the long training
with the initial training set of our active deep learning method
(the preprocessed Ondřejov dataset),
and we used the trained CNN to classify all the spectra in the LAMOST pool.
Then, we estimated precision of the CNN from random samples of 100 spectra from target classes.
In order to  make  a more reliable conclusion,
we ran the experiment three times.
The results are shown in~Table~\ref{tab:non-active}.

\begin{table}[h]
    \caption{
            Results of three runs of non-active learning.
            The table shows the precision estimated from a random sample of 100 spectra
            from each target class,
            and the numbers in brackets are counts of spectra classified into each target class.
    }
    \label{tab:non-active}
    \centering
    \begin{tabular}{c|cc}
        \hline\hline
        Run & \multicolumn{2}{c}{Estimated precision (predicted spectrum count)} \\
        & single peak & double peak \\
        \hline
            No. 1 & 4.1\% (343\,988) & 2.0\% (248\,336) \\
            No. 2 & 3.0\% (301\,396) & 2.0\% (409\,908) \\
            No. 3 & 4.0\% (167\,545) & 0.0\% (342\,230) \\
        \hline
    \end{tabular}
\end{table}

The comparison of~Table~\ref{tab:non-active} and Table~\ref{confusion} shows
that the three CNNs were unable to learn without the support of spectra from LAMOST
added to the training set by active learning.
Therefore we conclude that the gain of our active deep learning method is significant.

\section{Detailed analysis of emission-line candidates}
\label{appendix:reconfirmation}

All spectra in the candidate list predicted by our CNN were visually inspected to confirm the correct prediction of their classes.
We reviewed not only the limited spectral range around the H\(\alpha\) line given to the CNN
(which was only available in the Ondřejov dataset),
but also the whole LAMOST spectrum displayed in linear wavelengths
(instead of the original logarithmic scale of the input FITS files).

We removed from the candidate spectra those that were clearly bad
(reduction artefacts, missing data in the studied spectral range)
or were so noisy
that the line profile drawn as a continuous line was broken in a sawtooth-like manner.
Other bad spectra belonged to cold stars,
where the molecular bands, smeared because of the low resolution of LAMOST,
mimicked the searched emission profile in a small zoomed range,
but the rest of the continua had a similar variability amplitude.
The rest were spectra in the non-target class showing only absorption lines without signatures of emission.
We removed 58 spectra in total.
The list is available in the on-line tables published on Zenodo and CDS.
Despite being dropped in the uninteresting (non-target) class,
the visual inspection of apparently bad spectra also yielded several interesting objects, such
as high-velocity stars or objects with complicated physics as described in~Sects.~\ref{supernovae} and \ref{HVS}.

In general, it was challenging to decide about the quality of the data
and the correct classification in these boundary cases.
We took the size of the line profile disturbance,
the fuzziness of spectra (and namely its continua),
and other metadata (obtained from the Virtual Observatory as explained below)
into account to classify an object as the one with expected emission (e.g. a Be star).
Our overall experience with this very subjective classification in boundary cases was, however, very surprising.
The deep CNN was able to see even very tiny signatures of expected shape structures
that the human could barely see.

\subsection{Technology of the Virtual Observatory}
\label{appendix:votools}

The verification of the performance of our active deep learning method required many visualisations of spectra with the possibility of previewing entire spectra as well as their zoomed parts.
In many cases, we used the positional cross-matching followed by visual inspection of the appearance of a candidate object in common all-sky imaging and photometric surveys.
This task would be extremely tedious without the usage of the Virtual Observatory technology based on the IVOA standards,
namely the combination of the Table Access~\citep{TAP} and Simple Spectra Access~\citep{SSAP} protocols
and VO client applications such as TOPCAT~\citep{TOPCAT}, Aladin~\citep{ALADIN}, or SPLAT-VO~\citep{SPLAT-VO}.
All LAMOST DR2 FITS files, converted into the linear wavelength in \AA{}ngströms, were ingested into a VO server based on the DACHS system~\citep{DACHS} that runs locally in Ondřejov,
and the links (called \texttt{accref} or \texttt{access\_url}) to individual spectra on that server were joined with spectrum names in the candidate table.
This allowed an immediate visualisation and interactive zooming of every spectrum in SPLAT-VO.
We were unable to use the original China-VO services because the DR2 is not available via the SSAP protocol (only DR1 is available).

The combination of the TOPCAT, Aladin, and SPLAT-VO tools interlinked using the
SAMP protocol~\citep{SAMP} allowed us to set up a powerful workflow.  In
addition to basic operations on tables (e.g. sorting, counting, ordering, and
searching in rows) and cross-matching using internal TOPCAT capabilities, we extensively used the CDS cross-match service with SIMBAD running on CDS
computers.

The resulting tables were then sent to Aladin and SPLAT-VO, and the TOPCAT
activation actions were set on them, so that the selection of every row in the
TOPCAT table triggered a sending operation of the object coordinates to
Aladin.  Here the detailed image of a star or galaxy from the all-sky surveys
DSS2, 2MASS, SDSS, and GALEX quickly appeared thanks to the hierarchical
progressive survey (HiPS) technology~\citep{HIPS}.  In addition, the TOPCAT activation
action also triggered the sending of \texttt{accref} content to the
SPLAT-VO, where the corresponding spectrum was shown. We verified correct
cross-matching with SIMBAD by over-plotting all SIMBAD objects that are
visible in the field in Aladin (also based on the HIPS catalogues), and placing
the pointer at a particular target resulted in showing the SIMBAD web page
about the object in a browser.  More details are shown in our presentation
from the Paris IVOA interoperability meeting \citep{ivoa-paris}, which is available on
\href{https://doi.org/10.5281/zenodo.3242658}{Zenodo}\footnote{\url{https://doi.org/10.5281/zenodo.3242658}}.

The productivity of candidate verification was enormously increased by
the VO technology in comparison to a manual search for information in multiple
sources.  It also allowed us to discover the effects described below (e.g. misplacement of an optical fibre).  In addition to the VO
exploitation, SPLAT-VO was also used for direct measurement of radial
velocities through its built-in method of line profile
mirroring~\citep{skoda-parimucha}.  The vacuum wavelengths of the measured lines
were used as reference.

\subsection{Multiplicity of exposures}
\label{appendix:multiplicity}

Some objects in the LAMOST DR2 were observed several times,
usually twice in different epochs,
but a few spectra in our candidate list were also exposed five times.
This fact may be used for analysing the evolution of the line profiles, which is
typical for some Be stars.
For example, the comparison of spectrum \href{https://zenodo.org/record/3236166/files/spec-56625-GAC113N37V1\_sp07-098.pdf}{\small\textsf{spec-56625-GAC113N37V1\_sp07-098}}
of object LAMOST J074244.51+353401.3 exposed 281 days after the \href{https://zenodo.org/record/3236166/files/spec-56344-GAC113N37V1\_sp07-098.pdf}{\small\textsf{spec-56344-GAC113N37V1\_sp07-098}} shows a double-peak emission that diminishes in a deep absorption line,
while the spectrum \href{https://zenodo.org/record/3236166/files/spec-56350-GAC089N28V1\_sp04-121.pdf}{\small\textsf{spec-56350-GAC089N28V1\_sp04-121}}
of object LAMOST J055821.00+284549.6 exposed 474 days after the spectrum \href{https://zenodo.org/record/3236166/files/spec-55876-GAC\_089N28\_B2\_sp04-121.pdf}{\small\textsf{spec-55876-GAC\_089N28\_B2\_sp04-121}} shows the blue-shifted peak of a double-peak profile that decreases,
while the red-shifted peak is stable and the absorption in the line core is much deeper.

We grouped 855 of 4\,379 candidate spectra into 398 groups with the same designation as stated in the LAMOST header.
The remaining 3\,524 objects seemed to have only one exposure.
However, the internal cross-matching using the coordinates in the radius of 5\arcsec{}
and visual verification of spectra shape identified several objects with a different designation that lie close together.
We were able to identify 436 groups of multiple exposures within a radius of 5\arcsec{} .
However, this complicates the cross-matching because we cannot consider the object designation as a unique identifier.
For example, objects LAMOST J053611.80+273436.0 and LAMOST J053611.79+273435.9 are the same, as are LAMOST J041417.60+280609.6, LAMOST J041417.61+280609.5, and LAMOST J041417.62+280609.4.

We also found a case of two different stars with almost identical spectra: spectra \href{https://zenodo.org/record/3236166/files/spec-56204-GAC080N33B101\_sp08-234.pdf}{\small\textsf{spec-56204-GAC080N33B101\_sp08-234}}
of star \href{http://dr2.lamost.org/spectrum/view?obsid=59008234}{LAMOST J052402.81+334101.7}
and \href{https://zenodo.org/record/3236166/files/spec-56306-GAC080N33B2\_sp08-234.pdf}{\small\textsf{spec-56306-GAC080N33B2\_sp08-234}}
of star \href{http://dr2.lamost.org/spectrum/view?obsid=106508234}{LAMOST J052401.53+334120.9} exhibit almost identical H\(\alpha\) emission,
although the stars are 24\arcsec{} apart and have different SIMBAD names, \href{http://simbad.u-strasbg.fr/simbad/sim-id?Ident=2MASS+J05240280%2B3341017}{2MASS J05240280+3341017}
and \href{http://simbad.u-strasbg.fr/simbad/sim-id?Ident=2MASS+J05240153%2B3341210}{2MASS J05240153+3341210} .
We suppose that the two spectra are contaminated by the emission from the gaseous nebula around them,
or (see similar cases below) the spectrum of one star might be mixed with emission light of the second star before entering the fibre.

We still preserved the list of misclassified spectra
(58 in total, visually confirmed to be bad spectra or with an incorrect class, namely absorption) because they proved to be useful for the discovery of some peculiar objects.

\subsection{LAMOST classification pipeline flaws}

The LAMOST FITS headers contain the estimate of the spectral type of almost all stars
as well as labels produced by the automatic pipeline~\citep{pipeline} that mark the objects as non-stellar (galaxy, quasar, or unknown).
However, we noted various inconsistencies in these labels.
For example, object LAMOST J053040.90+260534.6 is classified in spectrum \href{https://zenodo.org/record/3236166/files/spec-56218-GAC083N27B2\_sp02-234.pdf}{\small\textsf{spec-56218-GAC083N27B2\_sp02-234}}
as star B6,
while in \href{https://zenodo.org/record/3236166/files/spec-56271-GAC084N26B1\_sp10-057.pdf}{\small\textsf{spec-56271-GAC084N26B1\_sp10-057}} it is A2V,
although the visually examined line profiles look very similar.
Although some stars are typical Be stars (identified by H16 as classical Be stars),
the LAMOST classification mostly assigns class A, but also class F or G.
Objects marked as `Non' are often just bad data with reduction artefacts,
but there are also some very interesting cases, as was shown in previous sections.

\subsection{Comparison with H16}

The main reason why we have used the LAMOST DR2 survey for our analysis instead of publicly available later versions
(e.g. DR4 with 7.6 million spectra is available since July 2018)
is not the lower data volume (which facilitates the whole analysis),
but the excellent opportunity of comparing our active deep learning methods with the more straightforward pattern-recognition algorithm of H16 on the same data set.
Their catalogue gives a more detailed line profile classification into six classes
(obtained by cross-correlation with 27 templates)
and attempts to cross-match the objects with other known emission-star catalogues, namely, with SIMBAD.

The detailed analysis of the cross-matching with our candidate list has, however, identified a number of problems
that prevented us from using their catalogue for a direct comparison of the performance of our method with theirs,
despite the same data set and the same target class (emission-line objects).

The first discrepancy is the non-unique identification of objects.
As they do not give spectrum identifications but only an object designation,
we joined our candidate list with their catalogue using the LAMOST designation
(which is in fact an encoding of coordinates in J2000).
However, the coordinate cross-matching using circles with a radius of 3\arcsec{} identified objects with coordinates stated by H16 that were different from those in the LAMOST header.
For example, for object LAMOST J015611.38+580928.6,
H16 gives a position that is 2.4\arcsec{} offset from those in the FITS header.
As the object J034031.33+504451.4 gives the largest coordinate discrepancy of 4.96\arcsec{} between coordinates in the H16 catalogue and the LAMOST header,
we cross-matched our candidates with H16 with a radius of 5\arcsec{},
instead of using just the verbatim match of designations.
We assume that the differences in coordinates stated in H16
and those we obtained from the FITS header may be caused by some post-processing of the LAMOST DR2 archive
before it was publicly released (H16 probably used the DR2 before its public release).

We also visually verified randomly selected spectra of objects from their catalogue
(using the multiple SSA query in TOPCAT at the given coordinates)
and realised that many objects in their catalogue were not justified to have the emission signatures in H\(\alpha\) line,
as the apparent emission bumps on line profiles were just a coincidence of a noise fluctuations,
similar to what we discuss at the beginning of Appendix~\ref{appendix:reconfirmation}.
This concerns most of the objects that are classified by them as unknown.
For example, the spectrum of \href{http://dr2.lamost.org/spectrum/view?obsid=200705070}{LAMOST J010450.45+423607.2} is completely spoiled by reduction artefacts that introduce a forest of narrow spikes everywhere.
Both spectra \href{http://dr2.lamost.org/spectrum/view?obsid=42607095}{\small\textsf{spec-55976-GAC\_084N40\_V1\_sp07-095}}
and \href{http://dr2.lamost.org/spectrum/view?obsid=83014033}{\small\textsf{spec-56257-GAC089N38V1\_sp14-033}}
of LAMOST J054458.54+382954.3 are noisy as well.

Many unknown types are probably not stars at all.
For example, object \href{http://dr2.lamost.org/spectrum/view?obsid=172711092}{LAMOST J000943.24+495009.2}
is, in fact, the galaxy \href{http://simbad.u-strasbg.fr/simbad/sim-coo?Coord=00+09+43.24000000+%2B49+50+09.2000000&Radius=2&Radius.unit=arcsec}{LEDA 2354919}.
Many objects of 1\,919 marked by H16 as unknown are therefore probably not well justified as emission-line stars.

The same problem concerns 3\,597 objects marked as the \ion{H}{II} region.
For example, all four available spectra
\href{http://dr2.lamost.org/spectrum/view?obsid=181506069}{\small\textsf{spec-56618-GAC057N34B1\_sp06-069}},
\href{http://dr2.lamost.org/spectrum/view?obsid=203503203}{\small\textsf{spec-56661-GAC061N34B1\_sp03-203}},
\href{http://dr2.lamost.org/spectrum/view?obsid=210506069}{\small\textsf{spec-56680-GAC057N34B2\_sp06-069,}}
and \href{http://dr2.lamost.org/spectrum/view?obsid=212403203}{\small\textsf{spec-56685-GAC061N34B2\_sp03-203}}
of object LAMOST J040124.35+343559.0 lack a signature of nebula emission lines,
and the apparent bump in the H\(\alpha\) core is just coincidence of noise
(compared with the noise amplitude in the continuum).

The situation is not better for 5\,580 stars marked by H16 as CBe type,  the classical Be stars.
Here, a random visual inspection identified a number of spectra that are too noisy to be able to assess the profile
or spectra without any emission signatures.
For example, the single-spectrum objects
\href{http://dr2.lamost.org/spectrum/view?obsid=182507092}{J000414.96+463022.6},
\href{http://dr2.lamost.org/spectrum/view?obsid=178105131}{J000955.98+392419.0},
\href{http://dr2.lamost.org/spectrum/view?obsid=59310225}{J003506.02+271339.9}
and \href{http://dr2.lamost.org/spectrum/view?obsid=61805222}{J004201.02+433802.0}
that were classified by H16 as CBe with H\(\alpha\) type II profile
or \href{http://dr2.lamost.org/spectrum/view?obsid=19703166}{J052630.97+291516.9}
with line profile classified as type VI are completely noisy.

For objects with multiple exposures, one spectrum is often extremely noisy
(but because of the noise fluctuation, the profile may be judged as having some tiny emission in absorption),
while the others show clearly only a pure absorption profile.
This seems to be the case of many CBe type II profiles
(e.g. spectra \href{http://dr2.lamost.org/spectrum/view?obsid=5103189}{\small\textsf{spec-55880-B8004\_3\_sp03-189}}
and \href{http://dr2.lamost.org/spectrum/view?obsid=25403192}{\small\textsf{spec-55930-B5593002\_sp03-192}}
of object J024156.88+535432.5)
or CBe type VI (e.g. spectra \href{http://dr2.lamost.org/spectrum/view?obsid=127804081}{\small\textsf{spec-56350-GAC089N28V2\_sp04-081}}
and \href{http://dr2.lamost.org/spectrum/view?obsid=212104081}{\small\textsf{spec-56684-GAC089N28V2\_sp04-081}}
of object J055836.34+283405.1,
spectra \href{http://dr2.lamost.org/spectrum/view?obsid=2414198}{\small\textsf{spec-55875-B7505\_sp14-198}}
and \href{http://dr2.lamost.org/spectrum/view?obsid=15314198}{\small\textsf{spec-55910-B91003\_sp14-198}}
of object J015840.60+583322.6,
or spectra \href{http://dr2.lamost.org/spectrum/view?obsid=4501140}{\small\textsf{spec-55879-B7905\_1\_sp01-140}}
and \href{http://dr2.lamost.org/spectrum/view?obsid=15301140}{\small\textsf{spec-55910-B91003\_sp01-140}}
of object J021505.52+565827.8).

We also found objects where all spectra are just extremely noisy.
For example, this is a case of spectra \href{http://dr2.lamost.org/spectrum/view?obsid=185606068}{\small\textsf{spec-56627-HD095359N274143M01\_sp06-068}}
and \href{http://dr2.lamost.org/spectrum/view?obsid=213810078}{\small\textsf{spec-56687-HD101242N281431M\_sp10-078}}
of object J100140.14+274030.6.

Taking all the uncertainties described above into consideration,
we were unable to estimate the performance of our method by a direct comparison with H16,
as many of their objects with reported emission are not well justified.
After some experiments, we had 2\,644 spectra that were identified by our active deep learning procedure
that were also justified to be discovered by H16.

\subsection{Cross-matching with SIMBAD}

As many objects in our candidate list look interesting because they resemble Be stars, cataclysmic variables, or young stellar objects,
it is important to find more information about them.
We therefore tried to cross-match them with the SIMBAD database using the CDS cross-match service built-in the TOPCAT.
It is based on finding the minimum angular distance between object coordinates
and the SIMBAD catalogue in a small circle of a given radius on the sky.
As object coordinates, we took the coordinates given in the LAMOST FITS headers,
which represent the coordinates on which the LAMOST optical fibre was placed.

The simple idea of cross-matching using tight tolerance (of the order
1--2\arcsec{}) appeared to be incorrect when
we started to verify the given position in the all-sky surveys DSS2, 2MASS, and SDSS (using the Aladin sky atlas).
We noted faint spectra of objects with a prominent emission profile on coordinates,
where there was no visible object in the sky surveys,
but there were bright stars nearby,
for which SIMBAD stated a young star or directly the emission-line object.
The spectra in some fibres probably came from the bright object at a distance of
even tens of arcseconds from the fibre position.
We therefore finally cross-matched our list of candidates with SIMBAD in different circles of sizes 5\arcsec{} up to 300\arcsec{}
and inspected the SIMBAD type of objects with larger distances
until we confirmed that the match was incorrect
(the nearest SIMBAD objects was not of emission nature).
The number of cross-matched objects also started to rise steeply after a certain tolerance was met,
indicating that the nearest object was not the correct one.
After some iterations, we set the acceptable radius for a SIMBAD cross-match to 20\arcsec{},
and we visually inspected
that the objects we cross-matched with SIMBAD were reasonably bright targets of emission type.
In several cases, we found a galaxy or a nebula
(and the corresponding cross-matched object was confirmed by viewing at its spectrum).

Here are few examples of misplaced light entering the fibre:
spectrum \href{https://zenodo.org/record/3236166/files/spec-56618-GAC105N47V2\_sp10-151.pdf}{\small\textsf{spec-56618-GAC105N47V2\_sp10-151}}
of object \href{http://dr2.lamost.org/spectrum/view?obsid=181710151}{LAMOST J065632.72+461614.9}
is the spectrum of star \href{http://simbad.u-strasbg.fr/simbad/sim-coo?Coord=104.13636+46.270819+&CooFrame=FK5&CooEpoch=2000&CooEqui=2000&CooDefinedFrames=none&Radius=14&Radius.unit=arcsec}{Psi~9~Aur}
offset by 13\arcsec{},
or \href{https://zenodo.org/record/3236166/files/spec-56295-VB056N24V1\_sp08-031.pdf}{\small\textsf{spec-56295-VB056N24V1\_sp08-031}}
of object \href{http://dr2.lamost.org/spectrum/view?obsid=100908031}{LAMOST J034912.80+240820.0}
belongs to the bright (5 mag) Be star \href{http://simbad.u-strasbg.fr/simbad/sim-id?Ident=%40676291&Name=*%20%2028%20Tau}{28~Tau
(Pleione)},
which is located 22\arcsec{} away from the fibre position.

The spectra with a prominent emission
\href{http://dr2.lamost.org/spectrum/view?obsid=37010152}{\small\textsf{spec-55960-GAC\_101N09\_V1\_sp10-152}},
\href{http://dr2.lamost.org/spectrum/view?obsid=37110152}{\small\textsf{spec-55960-GAC\_101N09\_V2\_sp10-152}} and
\href{http://dr2.lamost.org/spectrum/view?obsid=40110152}{\small\textsf{spec-55968-GAC\_101N09\_V3\_sp10-152}}
of object LAMOST J063910.49+084435.4 are probably emitted by nebula around the Herbig Ha/Be star \href{http://simbad.u-strasbg.fr/simbad/sim-id?Ident=%40907222&Name=V* R Mon}{R~Mon},
which lies 26\arcsec{} away.

\begin{figure}
    \resizebox{\hsize}{!}{
        \includegraphics{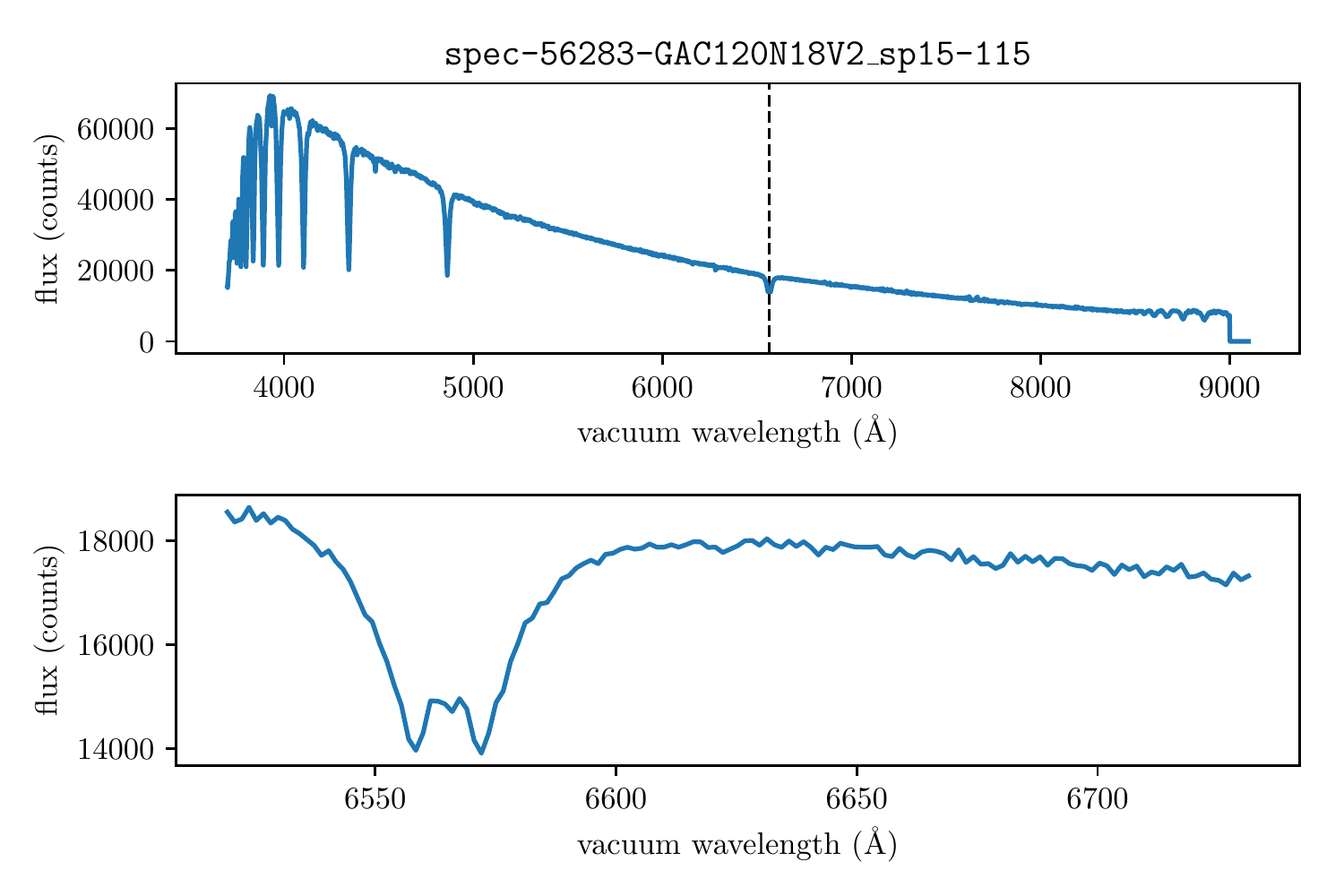}}
    \resizebox{\hsize}{!}{
        \includegraphics{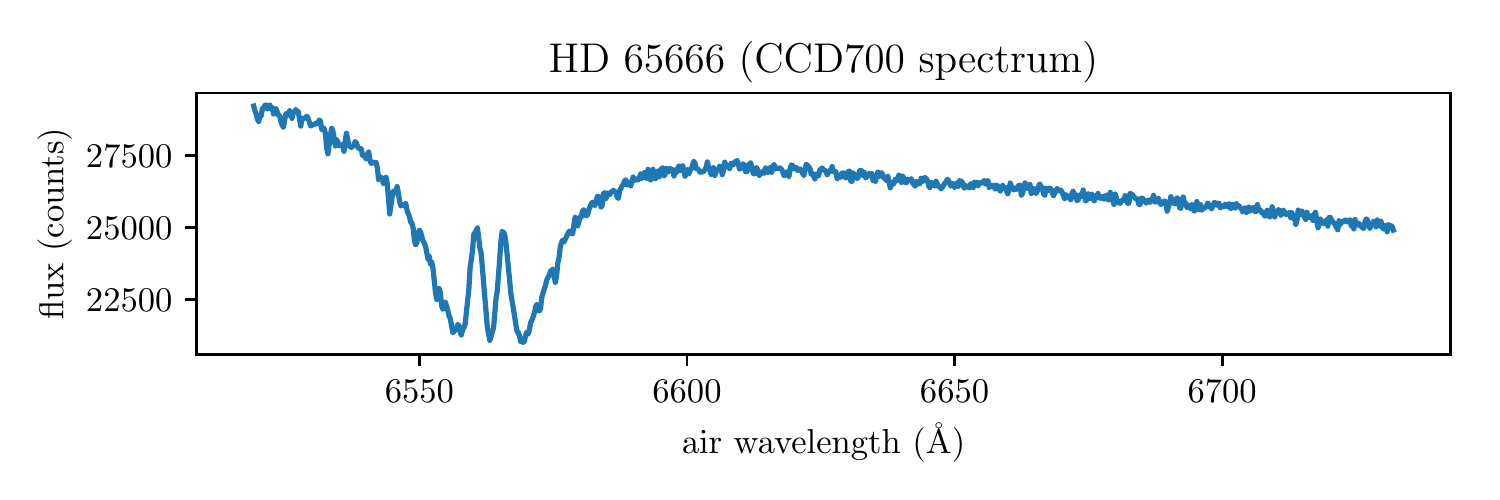}}
        \caption{
            Spectrum of the star that is cross-matched as HD~65\,666.
            The upper two panels are from LAMOST,
and             the lower panel is from the CCD700 archive of the Perek 2~m telescope.
            It is a bright star: \(V = 7.1\) mag.
            However, the LAMOST coordinates are offset by 6.5\arcsec{},
            so it is difficult to cross-match it with SIMBAD.
        }
    \label{hd65666}
\end{figure}

A very interesting case was also found for spectrum \href{https://zenodo.org/record/3236166/files/spec-56283-GAC120N18V2\_sp15-115.pdf}{\small\textsf{spec-56283-GAC120N18V2\_sp15-115}}
of object \href{http://dr2.lamost.org/spectrum/view?obsid=92715115}{LAMOST J080032.50+185028.9,}
which is marked by the LAMOST pipeline as an A1V star with \(r_{\mathrm{mag}}\) of value 11.04 (not identified in H16).
It is the bright seventh magnitude star \href{http://simbad.u-strasbg.fr/simbad/sim-coo?Coord=+120.13542+18.841371&CooFrame=FK5&CooEpoch=2000&CooEqui=2000&CooDefinedFrames=none&Radius=7&Radius.unit=arcsec}{HD~65\,666}
with a fibre offset by 6.5\arcsec{}.
Figure~\ref{hd65666} also shows its spectrum obtained by the Ondřejov 2~m Perek Telescope with a spectral resolution 13\,000,
which unveils a more complex structure of the line profiles that is typical for Be stars.

\end{appendix}
\end{document}